\documentclass[numberedappendix]{emulateapj}
\pdfoutput=1

\usepackage{apjfonts}
\usepackage{amsmath} 
\usepackage{color} 
\usepackage[colorlinks=true,linkcolor=blue,urlcolor=blue,citecolor=blue]{hyperref}

\shorttitle{The most massive stars in Arches and Quintuplet}
\shortauthors{F.R.N.~Schneider et al.}

\newcommand{\msun}{{\rm M}_{\odot}}
\newcommand{\lsun}{{\rm L}_{\odot}}
\newcommand{\rsun}{{\rm R}_{\odot}}

\bibpunct{(}{)}{;}{a}{}{,}

\begin{document}

\title{The ages of young star clusters, massive blue stragglers and the upper mass limit of stars:\\Analysing age dependent stellar mass functions}

\email{fschneid@astro.uni-bonn.de}
\author{
	F.R.N.~Schneider\altaffilmark{1},
	R.G.~Izzard\altaffilmark{1},
	S.E.~de~Mink\altaffilmark{2,3,*},
	N.~Langer\altaffilmark{1},
	A.~Stolte\altaffilmark{1},
	A.~de~Koter\altaffilmark{4,5},
	V.V.~Gvaramadze\altaffilmark{6,7},
	B.~Hu{\ss}mann\altaffilmark{1},
	A.~Liermann\altaffilmark{8,9} and
	H.~Sana\altaffilmark{4,10}
}
\altaffiltext{1}{Argelander-Institut f{\"u}r Astronomie der Universit{\"a}t Bonn, Auf dem H{\"u}gel 71, 53121 Bonn, Germany}
\altaffiltext{2}{Observatories of the Carnegie Institution for Science, 813 Santa Barbara St, Pasadena, CA 91101, USA}
\altaffiltext{3}{Cahill Center for Astrophysics, California Institute of Technology, Pasadena, CA 91125, USA}
\altaffiltext{4}{Astronomical Institute 'Anton Pannekoek', Amsterdam University, Science Park 904, 1098 XH, Amsterdam, The Netherlands}
\altaffiltext{5}{Instituut voor Sterrenkunde, KU Leuven, Celestijnenlaan 200D, 3001, Leuven, Belgium}
\altaffiltext{6}{Sternberg Astronomical Institute, Lomonosov Moscow State University, Universitetskij Pr. 13, Moscow 119992, Russia}
\altaffiltext{7}{Isaac Newton Institute of Chile, Moscow Branch, Universitetskij Pr. 13, Moscow 119992, Russia}
\altaffiltext{8}{Max Planck Institut f{\"u}r Radioastronomie, Auf dem H{\"u}gel 69, 53121 Bonn, Germany}
\altaffiltext{9}{Leibniz Institute for Astrophysics Potsdam (AIP), An der Sternwarte 16, 14482 Potsdam, Germany}
\altaffiltext{10}{Space Telescope Science Institute, 3700 San Martin Drive, Baltimore, MD 21218, USA}
\altaffiltext{*}{Einstein fellow}

\begin{abstract}
Massive stars rapidly change their masses through strong stellar winds and mass
transfer in binary systems. The latter aspect is important for
populations of massive stars as more than $70\%$ of all O-stars are expected to interact with
a binary companion during their lifetime.
We show that such mass changes leave characteristic signatures in stellar mass functions
of young star clusters which can be used to infer their ages and to 
identify products of binary evolution.
We model the observed present day mass functions of the young Galactic
Arches and Quintuplet star clusters using our rapid binary evolution code.
We find that shaping of the mass function by stellar wind mass loss allows us to
determine the cluster ages to $3.5\pm0.7\,\mathrm{Myr}$ and $4.8\pm1.1\,\mathrm{Myr}$,
respectively.
Exploiting the effects of binary mass exchange on the cluster mass function,
we find that the most massive stars in both clusters are rejuvenated 
products of binary mass transfer, i.e. the massive counterpart of classical 
blue straggler stars. This resolves the problem of an apparent age spread 
among the most luminous stars exceeding the expected duration of star formation 
in these clusters. We perform Monte Carlo simulations to probe stochastic sampling, which 
support the idea of the most massive stars being rejuvenated binary products.
We find that the most massive star is expected to be a binary 
product after $1.0\pm0.7\,\mathrm{Myr}$ in Arches and after $1.7\pm1.0\,\mathrm{Myr}$ 
in Quintuplet. Today, 
the most massive
$9\pm3$ stars in Arches and $8\pm3$ in Quintuplet are expected to be such objects. 
Our findings have strong implications for the stellar upper mass limit and
solve the discrepancy between the claimed $150\,\msun$ limit and observations of
fours stars with initial masses of $165$--$320\,\msun$ in R136 and of SN~2007bi, which is thought to be
a pair-instability supernova from an initial $250\,\msun$ star. Using the 
stellar population of R136, we revise the upper mass limit to values in the 
range $200$--$500\,\msun$.
\end{abstract}


\section{Introduction}\label{sec:introduction}
Massive stars play a key role in our Universe. They drive the 
chemical evolution of galaxies by synthesising most of the heavy 
elements. Their strong stellar winds, radiation feedback, powerful 
supernova explosions and long gamma ray bursts shape 
the interstellar medium. They are thought to
have played an essential role in reionising the Universe after 
the dark ages and are visible up to large distances.

Unfortunately, our understanding of the formation and evolution of the most massive
stars in the local Universe is incomplete \citep{2012ARA&A..50..107L}.
Recently it was established that most of the massive stars in 
the Milky Way are actually part of a binary star system 
and that more than $70\%$ of them will exchange mass 
with a companion during their life \citep{2012Sci...337..444S}.
Our understanding of these stars is further hampered by two major controversies.
The first one, the cluster age problem, 
concerns the ages of the youngest star clusters. 
Emerging star clusters are expected to form stars in a
time span shorter than the lifetime of their most massive
members \citep{2000ApJ...530..277E,2012ApJ...750L..44K}. 
In contrast, the most luminous
stars in two of the richest young clusters in our Galaxy, 
the Arches and Quintuplet clusters, show an apparently large age range. 
Their hydrogen- and nitrogen-rich Wolf-Rayet (WNh) stars appear significantly younger 
than most of their less luminous O stars 
\citep{2008A&A...478..219M,2012A&A...540A..14L}.
Similar age discrepancies are observed in other young stellar systems \citep{2003ARA&A..41...15M}
such as the Cygnus OB2 association
\citep{1999A&A...348..542H,2008A&A...485L..29G,2008A&A...487..575N}
and the star clusters Pismis~24 \citep{2011A&A...535A..29G} and NGC~6611
\citep{1993AJ....106.1906H,2008A&A...490.1071G}. 

The second controversy, the maximum stellar mass problem, concerns the stellar upper mass limit.
Such a upper mass limit is theoretically motivated by the Eddington-limit
which may prevent stellar mass growth by accretion above a certain mass \citep{1971A&A....13..190L}.
Observationally, an upper mass limit of about $150\,\msun$
is derived from the individual stellar mass distributions of the Arches and the R136 clusters
\citep{2004MNRAS.348..187W,2005Natur.434..192F,2006MNRAS.365..590K}
and from a broader analysis of young stellar clusters \citep{2005ApJ...620L..43O}.
This result is questioned by a recent analysis of very massive
stars in the core of R136, in which stars with initial masses of up to 
about $320\,\msun$ are found \citep{2010MNRAS.408..731C}.
Furthermore, recently detected ultra-luminous
supernovae in the local Universe are interpreted as explosions of
very massive stars \citep{2007A&A...475L..19L} 
--- e.g. SN~2007bi is well explained by a pair-instability
supernova from an initially $250\,\msun$ star \citep{2009Natur.462..624G,2009Natur.462..579L}.

Here, we show that both controversies can be resolved by considering a time dependent stellar mass 
function in young star clusters that accounts for stellar wind mass loss
and binary mass exchange. 
We perform detailed population synthesis calculations
of massive single and binary stars that include all relevant physical processes affecting the
stellar masses and compare them to observed present day mass functions of the Arches and Quintuplet
clusters. Our methods and the observations of the mass functions of the 
Arches and Quintuplet clusters are described in Sec.~\ref{sec:methods}. We
analyse the Arches and Quintuplet clusters in Sec.~\ref{sec:analysis-arches-quintuplet} to derive 
cluster ages and identify possible binary products by fitting our models
to the observed mass functions. Stochastic sampling effects are investigated 
in Sec.~\ref{sec:stochastic-sampling-effects} and the implications of our findings 
for the upper mass limit are explored in Sec.~\ref{sec:upper-mass-limit}.
We discuss our results in Sec.~\ref{sec:discussion} and give final conclusions 
in Sec.~\ref{sec:conclusions}.

\section{Methods and observational data}\label{sec:methods}
We analyse the Arches and Quintuplet clusters in two steps. 
First, we model their observed stellar mass functions to e.g. determine 
the initial mass function (IMF) slopes and the cluster ages. We set up a dense grid 
of single and binary stars, assign each stellar system in the grid a probability
of existence given the initial distribution functions 
(cf. Sec.~\ref{sec:init-distr-fcts}) and evolve the stars in time using our rapid binary evolution 
code described (cf. Sec.~\ref{sec:binaryc}). Present-day mass functions are then constructed
from the individual stellar masses at predefined ages. This
ensures that all the relevant physics like stellar wind mass loss
and binary mass exchange, which directly affects stellar masses,
is factored in our mass functions.

Second, we investigate stochastic sampling effects to e.g. compute
the probability that the most massive stars in the Arches and Quintuplet clusters are 
binary products. To that end, we randomly draw single and binary stars 
from initial distribution functions until the initial cluster masses
are reached and, again, evolve the drawn stellar systems with our rapid 
binary evolution code. The set-up of these Monte Carlo experiments is described in 
detail in Sec.~\ref{sec:mc-experiments}.

The initial distribution functions used in the above mentioned steps are summarised in Sec.~\ref{sec:init-distr-fcts}
and an overview of the observations of the Arches and Quintuplet clusters to which we compare our models is
given in Sec.~\ref{sec:observations}. We bin mass functions in a non-standard way
to compare them to observations --- the binning procedure is described in Sec.~\ref{sec:binning-procedure}.

\subsection{Rapid binary evolution code}\label{sec:binaryc}
The details of our population synthesis code are described in \citet{Schneider+2013a} 
and \citet{2013ApJ...764..166D}. Here, we briefly summarise the most 
important methods and assumptions that are used to derive our results.

We use a binary population code \citep{2004MNRAS.350..407I,2006A&A...460..565I,2009A&A...508.1359I}
to evolve single and binary stars and follow the evolution of the stellar masses and of other stellar properties 
as a function of time. Our code is based on a rapid binary evolution
code \citep{2002MNRAS.329..897H} which uses analytic functions \citep{2000MNRAS.315..543H}
fitted to stellar evolutionary models with convective core overshooting
\citep{1998MNRAS.298..525P} to model the evolution of single stars
across the whole Hertzsprung-Russell diagram. We use a metallicity of $Z=0.02$. 

Stellar wind mass loss \citep{1990A&A...231..134N} is applied to all stars with luminosities
larger than $4000\,\lsun$ \citep{2000MNRAS.315..543H}. 
The mass accretion rate during mass transfer is limited 
to the thermal rate of the accreting star \citep{2001A&A...369..939W}.
Binaries enter a contact phase and merge
if the mass ratio of accretor to donor is smaller than a critical value at the onset of Roche lobe overflow \citep{2013ApJ...764..166D}. 
When two main sequence stars merge,
we assume that $10\%$ of the total mass is lost and that $10\%$ of the envelope
mass is mixed with the convective core \citep{2013ApJ...764..166D}.

Photometric observations of star clusters 
cannot resolve individual binary components. In order to compare 
our models to observations we assume that binaries are unresolved in our models and determine masses 
from the combined luminosity of both binary components utilising our mass-luminosity relation.
Hence unresolved, pre-interaction binaries contribute to our mass functions.

We concentrate on main sequence (MS) stars because stars
typically spend about $90\%$ of their lifetime in this evolutionary
phase; moreover, our sample stars used for comparison are observationally colour-selected
to remove post-MS objects. If a binary is composed of a post-MS and a MS star, 
we take only the MS component into account. 

\subsection{Initial distribution functions for stellar masses and orbital periods}\label{sec:init-distr-fcts}
We assume that primary stars in binaries and single stars have masses $M_{1}$ distributed
according to a power law initial mass function (IMF) with slope $\Gamma$, 
\begin{eqnarray}
\xi(M_{1})=\frac{\mathrm{d}N}{\mathrm{d}M_{1}} & = & A\, M_{1}^{\Gamma -1}\,,\label{eq:gammadef}
\end{eqnarray}
in the mass range $1$ to $100\,\msun$ (where $A$ is a normalisation
constant). Secondary star masses, $M_{2}$, are taken from
a flat mass ratio distribution, i.e. all mass ratios $q=M_{2}/M_{1}\leq1$
are equally probable \citep{2012Sci...337..444S}. The initial orbital
periods for binaries with at least one O-star, i.e. a primary star
with $M_{1}\gtrsim15\,\msun$, mass ratio $q\geq0.1$ and a period
$0.15\leq\log (P/{\rm d}) \leq3.5$ are taken from the distribution 
of stars in Galactic open clusters \citep{2012Sci...337..444S}.
The initial periods of all other binaries follow a flat distribution
in the logarithm of the orbital period \citep{1924PTarO..25f...1O}.
Orbital periods are chosen such that all interacting binaries are
taken into account, i.e. the maximum initial orbital separation is
$10^{4}\,\rsun$ ($\approx 50\,{\rm AU}$). Binaries with wider orbits would be effectively single stars.

\subsection{Monte Carlo experiments}\label{sec:mc-experiments}
To address stochastic sampling, we perform Monte Carlo
simulations of star clusters and investigate the probability that
the most massive star in a star cluster is a product of binary evolution
as a function of cluster mass, age, binary fraction and IMF slope.

We assume that all stars are coeval and that every star cluster forms from a finite supply of mass 
with stellar masses stochastically sampled from initial
distribution functions. While single stars are sampled from the initial
mass function, binary stars are chosen from a larger parameter space
of primary and secondary masses and orbital periods. This larger parameter
space is better sampled in clusters of higher mass.

We draw single and binary stars
for a given binary fraction from the initial distribution functions
of primary mass, secondary mass and orbital period until a given initial
cluster mass, $M_{{\rm cl}}$, is reached. Here we consider
only stars with masses in the range
of $1$--$100\,\msun$. The true cluster masses are therefore 
larger if stars below $1\,\msun$ are added. Including these stars
according to a Kroupa IMF \citep{2001MNRAS.322..231K}, increases the
true cluster mass by $20\%$ and $89\%$ for high mass ($\geq 1\,\msun$) 
IMF slopes of $\Gamma=-0.70$ and $\Gamma=-1.35$, respectively.

After the stellar content of a cluster is drawn, we evolve the stars
in time to analyse whether the most massive stars at a given
cluster age result from close binary interaction. Repeating this
experiment $1000$ times provides
the probability that the most massive cluster star formed from binary
interactions, how long it takes on average until the most massive
star is a product of binary evolution, $\left\langle \tau_{{\rm B}}\right\rangle $,
and how many stars have on average a mass larger than that of the most massive cluster
star ($M_{\mathrm{S}}$) which did not accrete from a companion, $\left\langle N(M>M_{{\rm S}})\right\rangle $.
The most massive star that did not accrete from a companion can be a genuine 
single star or a star in a binary where binary mass transfer has not yet happened.
From here on we refer to this star as `the most massive genuine single star'.
We evolve and distribute stars as described in Secs.~\ref{sec:binaryc} and~\ref{sec:init-distr-fcts}.

To compare our Monte Carlo simulations with observations
of the Arches and Quintuplet clusters, we need to know the corresponding cluster masses $M_{\mathrm{cl}}$ in our Monte Carlo experiments.
We use IMF slopes of $\Gamma=-0.7$ as later determined in Sec.~\ref{sec:comparison-mass-functions} for both
clusters. The observations used for comparison \citep{2005ApJ...628L.113S,2012A&A...540A..57H} are complete for masses
$>10\,\msun$, corresponding to $234$ and $134$ stellar systems
in the Arches and Quintuplet and integrated masses of stars more massive
than $10\,\msun$ of $7200$ and $3100\,\msun$, respectively. 

Our best fitting Monte Carlo models of
the central regions of Arches and Quintuplet with primordial binary fractions of $100\%$ and $60\%$
have $225\pm10$ stellar systems with an integrated (initial) 
mass of $(7993\pm361)\,\msun$ and $136\pm10$ stellar systems with an integrated initial mass of 
$(3240\pm244)\,\msun$, respectively. These models 
correspond to initial cluster masses of $M_{{\rm cl}}\approx1.5\times10^{4}\,\msun$ 
and $M_{{\rm cl}}\approx0.9\times10^{4}\,\msun$, respectively, in stars with $1\leq M/\msun \leq 100$.

We assume that binaries are resolved in our Monte Carlo calculations, contrary
to when we model mass functions in order to compare to observed mass functions. 
This is because we make theoretical predictions and are thus interested in individual 
masses of all stars regardless of them being in a binary or not.

\subsection{Observations}\label{sec:observations}
The observed present-day mass functions of the Arches and Quintuplet
clusters were obtained from NAOS/CONICA (NACO) photometry at the VLT. 
The Arches cluster was observed in 2002 over a field
of view (FOV) of $27''$. The centre of the Quintuplet cluster was
imaged with NACO over a FOV of $40''$ 
in 2003 and 2008, which allowed the construction
of a membership source list from proper motions. Both data sets were
obtained in the $H$ ($\lambda_c = 1.66\,\mu\mathrm{m}$) and 
$K_s$ ($\lambda_c = 2.18\,\mu\mathrm{m}$) passbands. The colour information is used to 
remove likely blue foreground interlopers, red clump and giant 
stars towards the Galactic Center line of sight. Details can be found
in \citet{2005ApJ...628L.113S} for the Arches cluster and in \citet{2012A&A...540A..57H} 
for the Quintuplet cluster.

In the case of the Arches cluster, the known radial variation of the 
extinction is removed prior to individual mass determination
\citep{2002A&A...394..459S} employing the extinction law of \citet{1985ApJ...288..618R}. 
Masses are then derived from the $K_s$ magnitudes of each star
by comparison with a $2\,\mathrm{Myr}$ Geneva isochrone \citep{2001A&A...366..538L}. 
In the case of the Quintuplet, the better photometric performance
allowed all sources to be individually dereddened to a $4\,\mathrm{Myr}$ Padova
MS isochrone \citep[and references therein]{2008A&A...482..883M}
using the recently updated near-infrared extinction law towards 
the Galactic Center line of sight \citep{2009ApJ...696.1407N}. 
As detailed in \citet{2012A&A...540A..57H},
isochrone ages of $3$ and $5\,\mathrm{Myr}$ do not significantly alter the 
shape and slope of the constructed mass function. All mass 
determinations are based on solar metallicity evolution models.

With the aim to minimise any residual field
contamination, only the central $r < 10''$ or $0.4\,\mathrm{pc}$ of the Arches 
and $r < 12.5''$ or $0.5\,\mathrm{pc}$ of the Quintuplet (at an assumed
distance of $8.0\,\mathrm{kpc}$ to the Galactic Center; 
\citealt{2008ApJ...689.1044G}) were selected to 
construct the mass functions.
For the Arches cluster, this radial selection corresponds 
approximately to the half-mass radius, which implies that the 
mass projected into this annulus is of the order of $\sim 10^4\,\msun$
\citep[see][]{2009A&A...501..563E,2012ApJ...751..132C,2013A&A...556A..26H}.
In the Quintuplet cluster, the total mass is estimated to be
$6000\,\msun$ within the considered $0.5\,\mathrm{pc}$ radius \citep{2012A&A...540A..57H}.
The mass functions in the central regions of Arches and Quintuplet
have slopes that are flatter than the usual Salpeter slopes, most likely 
because of mass segregation \citep{2010MNRAS.409..628H,2013A&A...556A..26H}.

The most massive stars in the Arches and Quintuplet clusters
are hydrogen and nitrogen rich WNh stars. 
As reliable masses cannot be derived for these Wolf-Rayet (WR) stars from 
photometry alone, and as several of the WRs in the Quintuplet suffered
from saturation effects, the most massive stars are excluded from 
the mass functions. This affects $6$ WNhs stars with uncertain masses 
in Arches and $3$ (plus $7$ post-MS, 
carbon rich WR stars) in Quintuplet. These WNh stars 
are expected to contribute to the high mass tail of the Arches and Quintuplet
mass functions. 

\subsection{Binning procedure of mass functions}\label{sec:binning-procedure}
Following \citet{2005ApJ...628L.113S}, we employ a binning 
procedure that renders the observed mass functions
independent of the starting point of the bins. 
We shift the starting point by one tenth of the binsize and create
mass functions for each of these starting points.
We use a fixed binsize of $0.2\,{\rm dex}$ to ensure that the number of stars in each bin
is not too small and does not introduce a fitting bias \citep{2005ApJ...629..873M}. 
Each of these ten mass functions with different starting points is shown when 
we compare our mass functions to observations.

This procedure results in lowered number counts
in the highest mass bins because only the most massive stars will fall into
these bins as seen in the power-law mass function
(black dotted lines in Fig.~\ref{fig:observations}) where a kink
is visible around $\log M/\msun \approx 1.85$ (left panel) and $\log M/\msun \approx 1.45$ (right panel)
respectively (cf. convolution of a truncated horizontal 
line with a box function with the width of the bins). This kink is caused by
the binning procedure. Importantly, the observations, our models and the power-law mass functions
in Figs.~\ref{fig:observations} and \ref{fig:sfh} are binned identically to render
the mass functions comparable.

\section{Analyses of the Arches and Quintuplet clusters}\label{sec:analysis-arches-quintuplet}
For a meaningful comparison of the modelled with the observed 
mass functions, the star cluster and the observations thereof need to fulfil certain criteria.
They should be
\begin{itemize}
\item between $\sim2\,\mathrm{Myr}$ and $\sim10\,\mathrm{Myr}$ in age
such that the wind mass loss peak in the mass function is present \citep[see below and][]{Schneider+2013a},
\item massive enough such that the mass function samples the largest masses
\item homogeneously analysed, with a complete \emph{present-day} mass function above $\sim 10\,\msun$.
\end{itemize}
Both, the Arches and Quintuplet clusters fulfil all criteria and are therefore 
chosen for our analysis. 

Other possible star clusters, which can be analysed in principle, 
are the Galactic Center cluster, NGC~3603~YC, Westerlund~1
and R136
in the Large Magellanic Cloud. Trumpler~14 and Trumpler~16 in the
Galactic Carina nebula are not massive enough and rather an OB star association
with stars of different ages, respectively. 
For Westerlund~1 \citep{2013AJ....145...46L} and NGC~3603~YC \citep{2013ApJ...764...73P}, 
present-day mass functions were recently derived.
A brief inspection of these results shows that both clusters may be suitable for
an analysis as performed here for the Arches and Quintuplet clusters. We will 
investigate this further in the near future.
Possibly, NGC~3603~YC is too young such that its mass function is not
yet altered enough by stellar evolution to apply our analysis.

\subsection{The Arches and Quintuplet mass functions}\label{sec:comparison-mass-functions}

\begin{figure*}
\center
\includegraphics[width=0.9\textwidth]{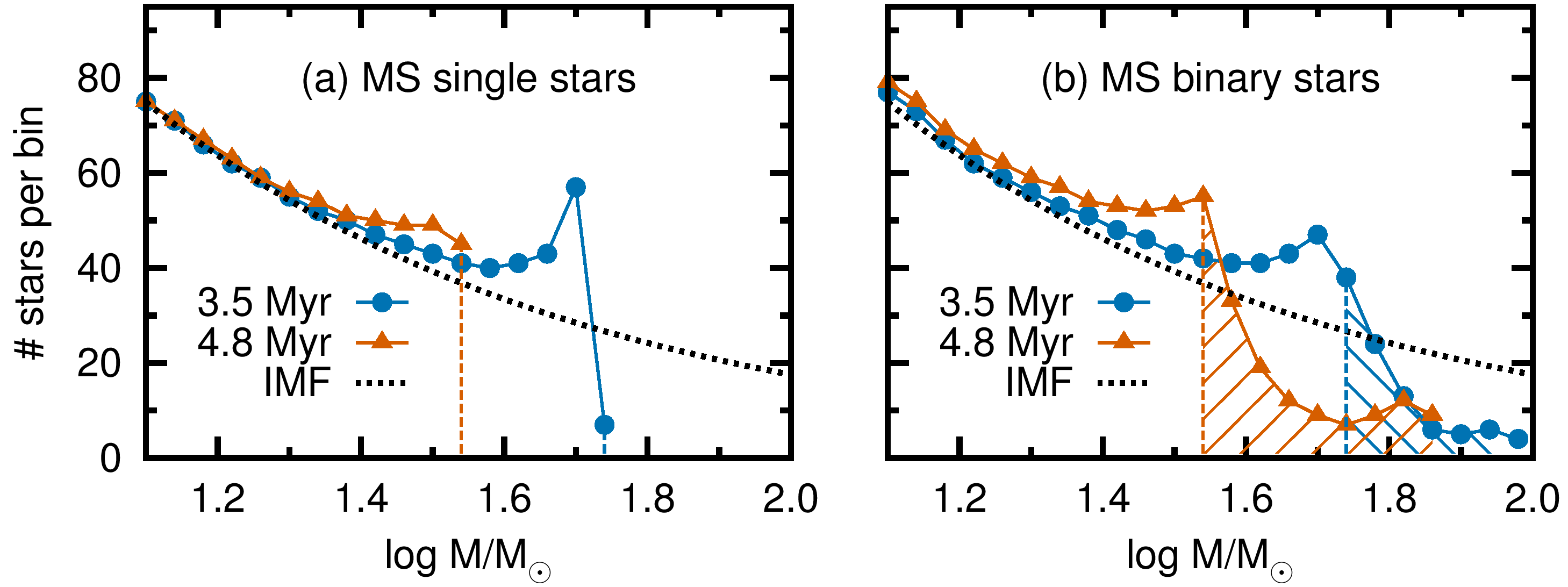}
\caption{Stellar mass functions, i.e. number of stars per logarithmic stellar 
mass bin, predicted by our population synthesis
models for main sequence (MS) single (panel a, left) and binary (panel
b, right) stars. Circles and triangles show the mass function at
$3.5$ and $4.8\,{\rm Myr}$ respectively. The black dotted line
shows the adopted initial mass function ($\Gamma =-0.7$). The peaks in the mass functions caused by
stellar wind mass loss are apparent in both plots at about $32\,\msun$ ($\log M/\msun\approx 1.5$) 
and $50\,\msun$ ($\log M/\msun\approx 1.7$) respectively.
The tail of stars affected by binary evolution in panel (b) is highlighted
by the hatched regions. The tail extends to about twice the maximum
mass expected from single star evolution, which is indicated by the vertical
dashed lines.}
\label{fig:models}
\end{figure*}

\begin{figure*}
\center
\includegraphics[width=0.9\textwidth]{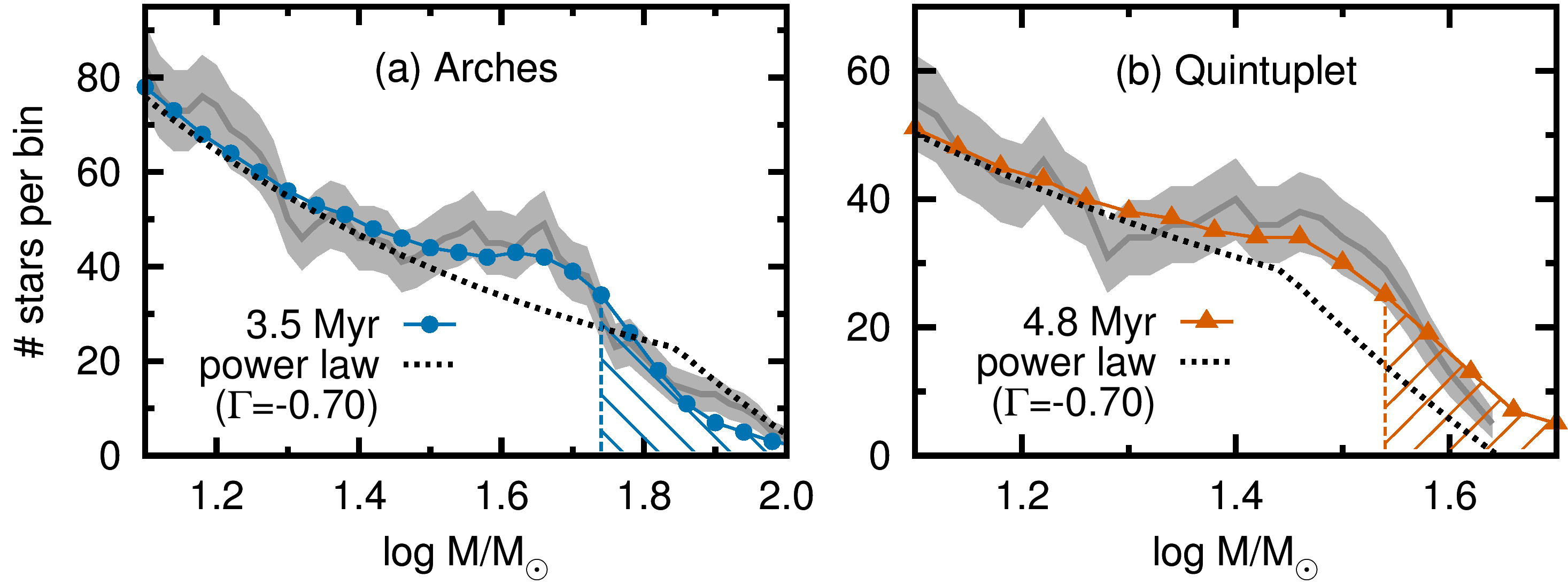}
\caption{Observed stellar mass functions for a binsize of $0.2\,\mathrm{dex}$ of (a) the Arches cluster \citep{2005ApJ...628L.113S} 
compared to our $3.5\,{\rm Myr}$ binary population model from Fig.~\ref{fig:models}b (primordial
binary fraction $100\%$) and (b) the Quintuplet cluster \citep{2012A&A...540A..57H} compared to our $4.8\,{\rm Myr}$
model (primordial binary fraction $60\%$). The peak and tail of the mass functions are well
reproduced by our models. Gray shaded regions
around the observed mass functions give Poisson uncertainties and show that the observed peaks deviate 
by $1$--$2\sigma$ from the power-law mass functions. 
We plot simple power-law mass functions as dotted lines, binned in the same way as
the observations and our models (Sec.~\ref{sec:binning-procedure}).
The kinks in the power-law mass functions result from our binning procedure.}
\label{fig:observations}
\end{figure*}

The initially most massive stars in a cluster end their life first. This depopulates the high mass
end of the stellar mass function. Before that, however, massive stars
lose a significant fraction of their initial mass because of strong stellar winds; 
e.g. our $100\,\msun$ star at solar metallicity loses about $40\,\msun$ during 
core hydrogen burning. Stellar wind mass loss shifts the top of the mass function towards lower masses
and a peak accumulates near its high mass end (Figs.~\ref{fig:models}a and \ref{fig:models}b). 
The location of the peak depends strongly on the cluster age and 
provides a clock to age-date a star cluster.

Stars in close binary systems exchange mass with their companion either by mass transfer or
in a stellar merger. \emph{A fraction} of stars gain mass, 
producing a tail at the high mass end of the mass function (hatched
regions in Fig.~\ref{fig:models}b) which extends beyond the
most massive single-stars (Fig.~\ref{fig:models}a) by up to a factor of about two.
The mass gainers \emph{appear} younger than genuine single stars because
their convectively mixed stellar core grows upon mass accretion and mixes fresh
fuel into their centre, thereby turning their clock backwards \citep{1998A&A...334...21V}.
Furthermore, the most massive gainers reach masses which, when interpreted as single stars, 
have lifetimes that are shorter than the cluster age --- they are the massive counterpart
of classical blue straggler stars \citep{Schneider+2013a}.

The mass functions of the cores of the Arches ($r\lesssim 0.4\,{\rm pc}$) and 
Quintuplet ($r\lesssim 0.5\,{\rm pc}$) clusters
\citep{2005ApJ...628L.113S,2012A&A...540A..57H}
reveal both the stellar wind mass loss peak
and the tail because of binary mass exchange.
Compared to a power-law, we find that the Arches and Quintuplet mass functions
are overpopulated in the ranges $32$ to $50\,\msun$ ($\log M/\msun=1.5$--$1.7$)
and $20$ to $32\,\msun$ ($\log M/\msun=1.3$--$1.5$), respectively (Fig.~\ref{fig:observations}).

These peaks are well reproduced by our models 
(Figs.~\ref{fig:models} and \ref{fig:observations}).
We can thus determine the cluster age because among the stars in the peak
are the initially most massive stars that are (a) still on but about
to leave the main sequence and (b) unaffected by binary interactions 
(we refer to them as turn-off stars in analogy to their position 
close to the turn-off in a Hertzsprung-Russell diagram).
The majority of stars in the peak
are turn-off stars but there are small contributions from
unresolved and post-interaction binaries \citep[see][]{Schneider+2013a}.
A correction for wind mass loss then reveals the initial mass of 
the turn-off stars and hence the age of the cluster. 

We can correct for wind mass loss by redistributing the number of excess stars in the peak $N$ 
such that the mass function is homogeneously filled for masses larger than those of the peak stars 
up to a maximum mass, the initial mass of the turn-off stars $M_\mathrm{to,i}$. 
The number of excess stars is then
\begin{equation}
N = \int_{M_\mathrm{to,p}}^{M_\mathrm{to,i}} \, \xi(M)\,\mathrm{d}M \label{eq:number-excess-stars}
\end{equation}
where $M_\mathrm{to,p}$ is the present-day mass of the turn-off stars, 
which can be directly read-off from the upper mass end of the peak, and $\xi(M)$ 
the initial mass function as defined in Eq.~(\ref{eq:gammadef}). The initial mass of the turn-off
stars $M_\mathrm{to,i}$ and hence the cluster age follows from integrating 
Eq.~(\ref{eq:number-excess-stars}),
\begin{equation}
M_\mathrm{to,i} = \left( \frac{N\Gamma}{A} + M_\mathrm{to,p}^\Gamma \right)^{1/\Gamma}. \label{eq:turn-off-mass}
\end{equation}
The normalisations, $A$, of the mass functions to be filled up with the excess stars, $N$,
(dotted, power-law functions in Fig.~\ref{fig:observations}) are $A=964$ and $A=639$ for 
Arches and Quintuplet, respectively, with slopes of $\Gamma=-0.7$ in both cases 
(see discussion below for why the mass functions are so flat).
It is difficult to read-off the exact value of $M_\mathrm{to,p}$ from the
observed mass functions because of the binning. But from binning our modelled mass functions
in the same way as the observations, we know that $M_\mathrm{to,p}$ corresponds 
to the mass shortly after the peak reached its local maximum (cf. the 
vertical dashed lines in Figs.~\ref{fig:models} and~\ref{fig:observations}).
Depending on the exact value of $M_\mathrm{to,p}$, $\log M_\mathrm{to,p}/\msun = 1.70$--$1.74$ 
in Arches and $\log M_\mathrm{to,p}/\msun = 1.50$--$1.54$ in Quintuplet, we
find $12$--$14$ and $7$--$10$ excess stars in the peaks of the Arches and
Quintuplet mass functions, respectively. These numbers of excess stars result in turn-off masses 
$M_\mathrm{to,i}$ of $62$--$72\,\msun$ and $36$--$43\,\msun$ 
and hence ages of $3.8$--$3.5\,\mathrm{Myr}$ and $5.2$--$4.7\,\mathrm{Myr}$
for the Arches and Quintuplet clusters, respectively. These are only
first, rough age estimates that will be refined below and their 
ranges stem from the uncertainty in reading-off $M_\mathrm{to,p}$ from the observed mass functions.

From the difference between the initial and present day masses of the
turn-off stars in Arches and Quintuplet, we can 
directly measure the amount of mass lost by these stars through stellar winds.
The turn-off stars in Arches lost about $12$--$17\,\msun$ and
the turn-off stars in Quintuplet about $4$--$8\,\msun$ during their
MS evolution. This is a new method to measure stellar wind mass loss 
which does not require measurements of stellar wind mass loss rates
and can therefore be used to constrain these.

More accurately, we determine the ages of the Arches and Quintuplet clusters
by fitting our population synthesis models (Sec.~\ref{sec:methods}) to the observed mass functions.
First, we fit power-law functions to the observed mass functions in mass 
regimes in which they are observationally complete and not influenced by stellar wind mass loss 
($10\lesssim M/\msun \lesssim 32$ and $10\lesssim M/\msun \lesssim20$, respectively). 
Binary effects are also negligible because stars with such masses 
are essentially unevolved at the present cluster ages. This fit gives the normalisation and a first 
estimate of the slope of the mass function. We then vary the mass function slope, 
the cluster age and the primordial binary fraction in our models simultaneously such
that the least-square deviation from the observations is minimised.
Our best-fit models are shown in Fig.~\ref{fig:observations} together with the observed mass functions.

We find slopes of $\Gamma=-0.7$, ages of $3.5\pm0.7\,{\rm Myr}$ and 
$4.8\pm1.1\,{\rm Myr}$ and primordial binary fractions of $100\%$
and $60\%$ for the Arches and Quintuplet cluster, respectively.
The binary fractions are less robust and may be the same within uncertainties 
because we do not take the uncertain masses of the WNh stars into account.
While our mass function fits contribute to the age uncertainties by only $\pm0.3\,\mathrm{Myr}$,
its major part, $\pm0.6\,\mathrm{Myr}$ and $\pm1.1\,\mathrm{Myr}$ for Arches and
Quintuplet, respectively, is due to observational uncertainties in stellar masses 
of $\pm30\%$ (Sec.~\ref{sec:discussion}).

Massive stars tend to sink towards the cluster cores because of dynamical friction 
(mass segregation), thereby flattening the mass function of stars in the core.
The derived mass function slopes of $\Gamma=-0.7$ are flatter than 
the typical Salpeter slope of $\Gamma=-1.35$ \citep{1955ApJ...121..161S} because we investigate only the mass 
segregated central regions of both clusters \citep[see][as well as Sec.~\ref{sec:dynamical-effects}]{2013A&A...556A..26H},
i.e. a subsample of stars biased towards larger masses.

In our models (Fig.~\ref{fig:models}), the tail of the Arches mass function contains about $30\%$ unresolved, 
pre-interaction binaries with $\log M/\msun \geq 1.76$ ($M \approx 58\,\msun$) and 
about $20\%$ with $\log M/\msun \geq 1.80$ ($M\approx 63\,\msun$). For Quintuplet, 
the fraction of unresolved, pre-interaction binaries is about $20\%$ 
with $\log M/\msun \geq 1.56$ ($M \approx 36\,\msun$) and 
about $10\%$ with $\log M/\msun \geq 1.60$ ($M\approx 40\,\msun$).
The binary fraction among the rejuvenated binary products in the tails is about 
$55\%$ in our Arches and $70\%$ in our Quintuplet model, where 
the remaining stars are single star binary products, i.e. merger stars.

\subsection{The ages of Arches and Quintuplet}\label{sec:ages-arches-and-quintuplet}
Previously estimated ages for the Arches and Quintuplet clusters lie in the range $2$--$4.5\,{\rm Myr}$
\citep{2001AJ....122.1875B,2002ApJ...581..258F,2008A&A...478..219M} and $2$--$5\,{\rm Myr}$
\citep{1999ApJ...514..202F,2010A&A...524A..82L,2012A&A...540A..14L}, respectively.
Within these ranges, the age discrepancy between the most luminous cluster members,
the WN and the less luminous O stars, accounts for about $1\,\mathrm{Myr}$ and $1.5\,\mathrm{Myr}$, 
respectively \citep{2008A&A...478..219M,2012A&A...540A..14L}, which is eliminated by our method. 
Our ages of $3.5\pm0.7\,{\rm Myr}$ and $4.8\pm1.1\,{\rm Myr}$ for the Arches and Quintuplet clusters, respectively, 
agree with the ages derived from the O~stars and dismiss the 
proposed younger ages from the brightest stars as a result of neglecting binary interactions.
The most famous member of the Quintuplet, the Pistol star, is such an example because
it appears to be younger than $2.1\,{\rm Myr}$ assuming single-star evolution \citep{1998ApJ...506..384F}.

\section{Stochastic sampling of binary populations}\label{sec:stochastic-sampling-effects}
The initial mass of the primary star, the mass ratio and the orbital 
period of a binary system determine when mass transfer starts, with more massive 
and/or closer binaries interacting earlier. Stochastic effects caused by the limited
stellar mass budget prevent the formation of all possible binaries in a stellar cluster,
i.e. binaries with all possible combinations of primary mass, mass ratio and orbital period.
The likelihood that a binary in a given cluster interacts, e.g. after $2\,\mathrm{Myr}$, 
and that the binary product becomes then the most massive star depends thus on 
the number of binary stars in that cluster, hence on the total cluster mass.
Using Monte Carlo simulations, we investigate 
the influence of stochastic sampling and binary evolution
on the most massive stars in young star clusters (cf. Sec.~\ref{sec:mc-experiments}).
 
The Galactic star cluster NGC~3603YC contains NGC~3603-A1, a binary star
with component masses $(116\pm31)\,\msun$ and $(89\pm16)\,\msun$
in a $3.77\,\mathrm{d}$ orbit \citep{2008MNRAS.389L..38S}. An initially
$120+90\,\msun$ binary in a $3.77\,\mathrm{d}$ orbit starts
mass transfer $\sim1.4\,\mathrm{Myr}$ after its birth
according to the non-rotating models of \citet{2012A&A...537A.146E}. This is the time needed for the 
$120\,\msun$ primary star to fill its Roche lobe as a result of stellar evolutionary expansion. 
This time provides an upper age estimate for NGC~3603~YC.
After mass transfer, the secondary star will be the most massive star in the
cluster. Were NGC~3603-A1 in a closer orbit, it could already be
a binary product today.

To find the probability that the most massive star in a cluster of
a given age is a binary product, we investigate how many close binaries
are massive enough to become the most massive star by mass transfer. 
Were the cluster a perfect representation of the initial stellar distribution functions,
we could use these functions to derive the probability directly. However,
the finite cluster mass and hence sampling density must be considered
for comparison with real clusters. Returning to the example of NGC~3603~YC,
were the cluster of larger total mass, its binary parameter space
would be better sampled and its most massive star might already be
a binary interaction product. For perfect sampling, i.e. infinite 
cluster mass, the time until a binary product is the most massive star
tends towards zero.

The idea that the most massive star in a star cluster may be a binary
product resulted from the first discovery of blue straggler stars
\citep{1953AJ.....58...61S}. It was proposed that blue stragglers might stem from
binary mass transfer and/or stellar collisions \citep{1964MNRAS.128..147M,1976ApL....17...87H}.
Stellar population synthesis computations including binary stars then showed
that this is indeed possible \citep[e.g.][]{1984MNRAS.211..391C,1994A&A...288..475P,1998A&A...334...21V,2001MNRAS.323..630H,2009MNRAS.395.1822C}.
Here, we show --- using the binary distribution functions of \citet{2012Sci...337..444S} ---
that the formation of blue stragglers by binary interactions prevails 
up to the youngest and most massive clusters and quantify it for the Arches and Quintuplet clusters.

\begin{figure}
\center
\includegraphics[width=0.46\textwidth]{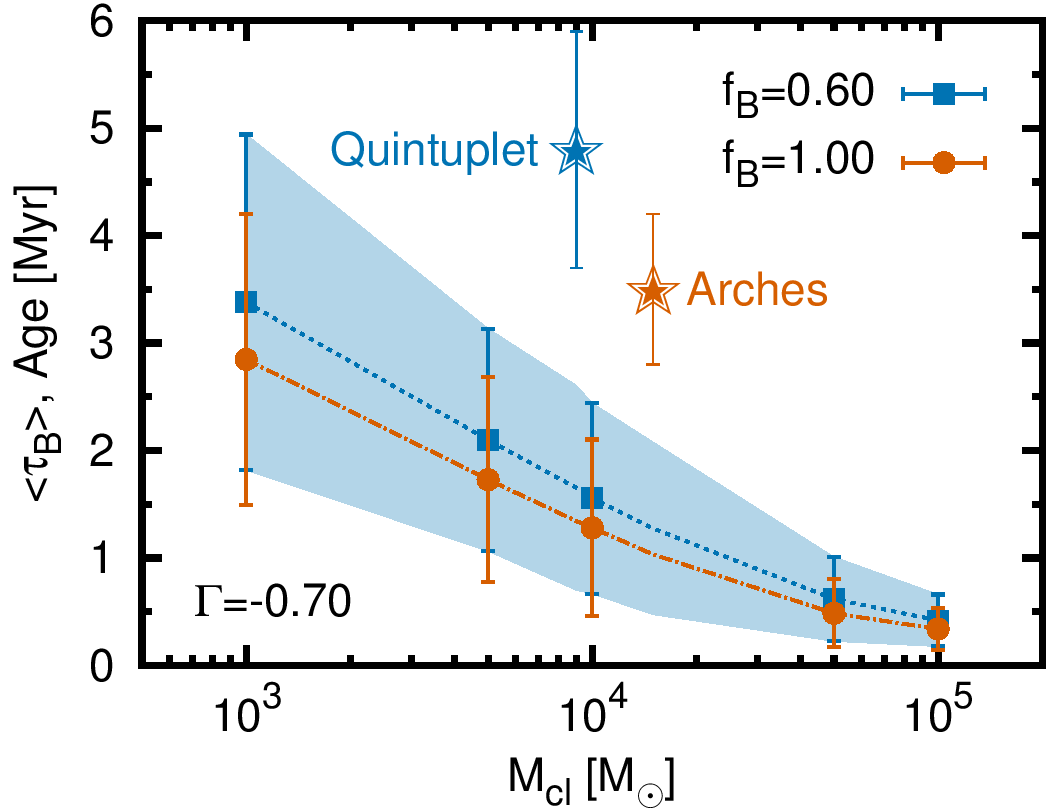}
\caption{Average time $\left\langle \tau_{{\rm B}}\right\rangle$ until the
most massive star in a star cluster is a product of close
binary evolution as a function of cluster mass for
two primordial binary fractions, $f_{\rm B}$, 
and a mass function slope $\Gamma=-0.7$. 
For steeper, Salpeter-like mass functions see Fig.~\ref{fig:average-times-until-interaction-2}.
The error bars are the standard deviation of $1000$ realisations of each cluster.
The star symbols indicate the age and \emph{central} cluster mass of Arches and Quintuplet
as derived in this work.}
\label{fig:average-times-until-interaction}
\end{figure}

In Fig.~\ref{fig:average-times-until-interaction},
we show the average time $\left\langle \tau_{{\rm B}}\right\rangle $
after which the most massive star in a star cluster is a product of
binary evolution as a function of the cluster mass $M_{{\rm cl}}$
for two different primordial binary fractions $f_{{\rm B}}$.
The error bars are $1\sigma$ standard deviations of $1000$
Monte Carlo realisations.
The slope of the mass function is $\Gamma=-0.7$, appropriate
for the mass-segregated central regions of both Arches and Quintuplet. The more massive
a star cluster, i.e. the more stars populate the multidimensional
binary parameter space, the shorter is this average time because the
probability for systems which interact early in their evolution
is increased. For less massive clusters $\left\langle \tau_{{\rm B}}\right\rangle $
increases and the statistical uncertainty grows.
For example if $M_{{\rm cl}}=10^{3}\,\msun$,
there are only about $16\pm3$ binaries in which at least one star
has a mass above $10\,\msun$ (for $\Gamma=-0.7$ and $f_{{\rm B}}=100\%$).
The same reasoning holds for different binary fractions: the higher
the binary fraction, the more binaries and hence a shorter
average time until the most massive star results from binary
interactions. 

\begin{figure}
\center
\includegraphics[width=0.46\textwidth]{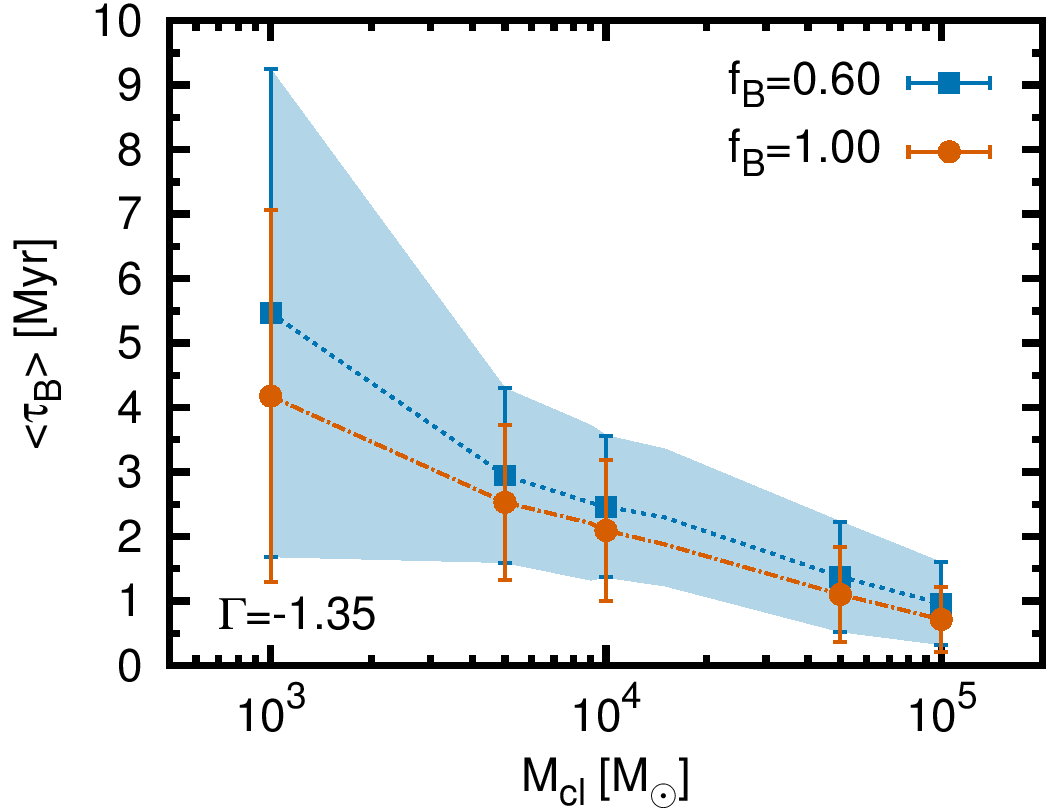}
\caption{As Fig.~\ref{fig:average-times-until-interaction} but for a steeper mass function
with a Salpeter slope of $\Gamma=-1.35$. The binary parameter space spanned by the initial mass ratios and 
initial orbital separations for massive primary stars is now less populated, resulting in increased
average times until the most massive star is a binary product. Similarly, the standard deviations increase.}
\label{fig:average-times-until-interaction-2}
\end{figure}

With a Salpeter mass function \citep[$\Gamma=-1.35$, ][]{1955ApJ...121..161S}
the average time until the most massive star is a binary product increases
compared to $\Gamma=-0.7$ (Fig.~\ref{fig:average-times-until-interaction-2})
because there are fewer massive binaries that interact to form the
most massive star. 
Assume a $4\,\mathrm{Myr}$ old star cluster has a mass function 
slope of $\Gamma=-1.35$, a total mass in stars above $1\,\msun$ of 
$M_\mathrm{cl}=10^4\,\msun$ (i.e. a true cluster mass of 
$1.9\times10^4\,\msun$ if stars below $1\,\msun$ follow a Kroupa IMF; see 
Sec.~\ref{sec:mc-experiments}) and a primordial binary fraction of $f_\mathrm{B}=60\%$. 
From Fig.~\ref{fig:average-times-until-interaction-2}, we can then read-off
after which time the most massive star is expected to be a binary product, 
namely after $2.5\pm1.1\,\mathrm{Myr}$.

The central regions of the Arches and Quintuplet clusters have
masses of $M_{{\rm cl}}=1.5\times10^{4}\,\msun$ and $0.9\times10^{4}\,\msun$ 
in stars more massive than $1\,\msun$ (Sec.~\ref{sec:mc-experiments}) 
and ages of $3.5\pm0.7$ and $4.8\pm1.1\,{\rm Myr}$
(Sec.~\ref{sec:comparison-mass-functions}), respectively. 
From Fig.~\ref{fig:average-times-until-interaction},
we expect that the most massive star in the Arches cluster is a binary
product after $1.0\pm0.7\,{\rm Myr}$ and after $1.7\pm1.0\,{\rm Myr}$
in the Quintuplet cluster. 

\begin{figure*}
\center
\includegraphics[width=0.7\textwidth]{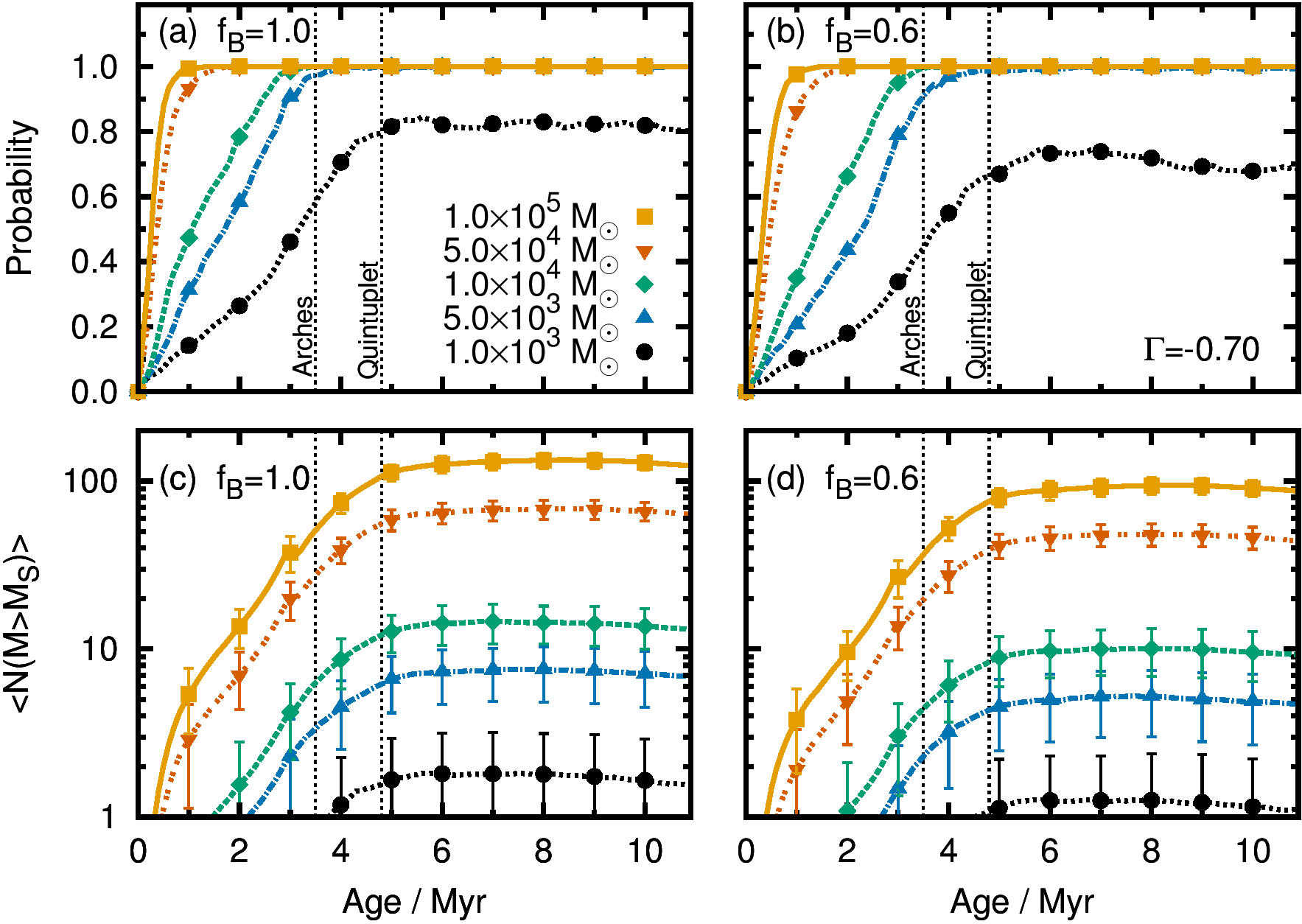}
\caption{Probability that the most massive star in a young star cluster is a product of binary evolution
(panels a~\&~b) and average number of stars more massive than the most
massive genuine single star $\left\langle N(M>M_{{\rm S}})\right\rangle $
(panels c~\&~d) as a function of age for several cluster masses $M_{{\rm cl}}$.
The left panels (a) and (c) have a binary fraction of $100\%$
whereas the right panels (b) and (d) have a binary fraction of
$60\%$. The symbols represent different cluster masses
$M_{{\rm cl}}$ and the error bars correspond to the $1\sigma$ standard deviation
of $1000$ realisations per cluster mass. The vertical dashed lines 
indicate the ages of the Arches and the Quintuplet clusters. 
The adopted IMF slope is $\Gamma=-0.7$.}
\label{fig:prob-numbers}
\end{figure*}

\begin{figure*}
\center
\includegraphics[width=0.7\textwidth]{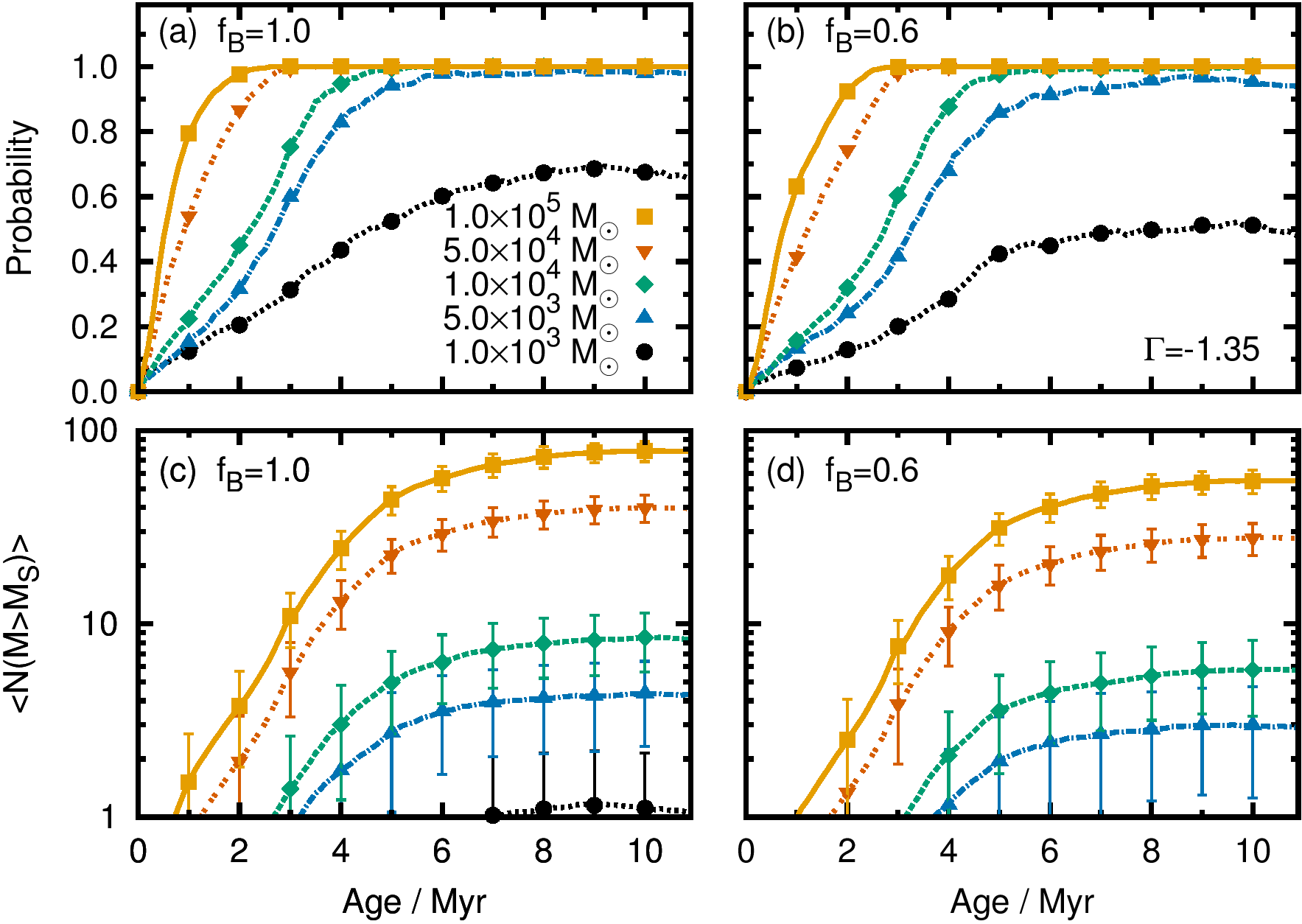}
\caption{As Fig.~\ref{fig:prob-numbers} with a Salpeter mass function, i.e.
slope of $\Gamma=-1.35$. The relative fraction of massive stars in each cluster
is reduced compared to $\Gamma=-0.7$.}
\label{fig:prob-numbers-2}
\end{figure*}

In Fig.~\ref{fig:prob-numbers} we show the probability that the most massive star
is a binary product and the average number of stars that are more massive
than the most massive genuine single star as a function of cluster age 
for different cluster masses and two different binary fractions. The IMF slope is
$\Gamma=-0.7$. The corresponding probabilities and average numbers 
for a Salpeter ($\Gamma=-1.35$) mass function are shown in Fig.~\ref{fig:prob-numbers-2}. 
Again, the error bars are $1\sigma$ standard deviations from $1000$ 
Monte Carlo experiments. Returning to the above mentioned example star cluster ($M_\mathrm{cl}=10^4\,\msun$): 
from Fig.~\ref{fig:prob-numbers-2}, we find that the most massive
star is a binary product with a probability of $88\%$ and that
the most massive $2.1\pm1.4$ stars are expected to be binary products
for the exemplary cluster age of $4\,\mathrm{Myr}$.

Given the ages of the Arches and Quintuplet clusters, we find
a probability of $>99.9\%$ that the most massive star
in each cluster is a binary product,
with the most massive $9.2\pm3.0$ and $7.5\pm2.8$ stars
being products of binary evolution in Arches and Quintuplet, respectively.
This is compatible with the number of WNh stars in the cores of Arches and Quintuplet,
which are the most luminous and hence most massive stars in these clusters, 
implying they are massive blue stragglers.

\section{The stellar upper mass limit}\label{sec:upper-mass-limit}
Data from two star clusters provide the current evidence for the existence 
of an upper stellar mass limit around $150\,\msun$: the Arches cluster in the 
Galactic Center \citep{2005Natur.434..192F} and the R136 cluster in the Large Magellanic Cloud 
\citep{2004MNRAS.348..187W,2005ApJ...620L..43O,2006MNRAS.365..590K}. However, according to our analysis
an upper mass limit cannot be derived from the Arches cluster because 
(a) it is too old, hence the most massive stars already exploded, and 
(b) its present-day high mass star population is dominated by binary products.
The situation might be different in the R136 cluster: current age estimates lie in the range
$1$--$4\,{\rm Myr}$ \citep{1995ApJ...448..179H,1998ApJ...493..180M,1998ApJ...509..879D,2009ApJ...707.1347A,2010MNRAS.408..731C}.
In the following we assume that the cluster is young enough such that even the most
massive stars have not yet evolved off the main sequence, to explore what we can learn from
R136 about a possible stellar upper mass limit.

Four stars in R136 with initial masses of $165$--$320\,\msun$
appear to exceed the currently discussed upper mass limit of $150\,\msun$  \citep{2010MNRAS.408..731C}. 
Either these stars were born with masses
exceeding $150\,\msun$ or gained mass
from other stars --- e.g. by binary interactions (this work) or dynamically induced
stellar mergers \citep{1999A&A...348..117P,2012MNRAS.426.1416B}.

From our Monte Carlo simulations (Sec.~\ref{sec:stochastic-sampling-effects}),
we cannot judge with high enough confidence whether the most massive star in 
R136 is expected to be a binary product or not because of the uncertain age of R136.
R136 has an IMF with approximately a Salpeter slope $\Gamma=-1.35$ \citep{1998ApJ...493..180M} and 
its cluster mass is $5$--$10\times 10^4\,\msun$ \citep{1995ApJ...448..179H,2009ApJ...707.1347A,2012A&A...546A..73H}.
From our Monte Carlo simulations of star clusters with binary fractions of 
$60\%$ and cluster masses $M_\mathrm{cl}$ of $5\times10^4\,\msun$ and $10^5\,\msun$ 
(Fig.~\ref{fig:prob-numbers-2}), we find that the most massive star is expected to be a 
binary product after $1\,\mathrm{Myr}$ with a probability of 
$42\%$ and $63\%$, respectively. The probabilities increase to $74\%$ and $92\%$, 
respectively, for a cluster age of $2\,\mathrm{Myr}$ and are larger than $98\%$ for
an age of $3\,\mathrm{Myr}$. So if the cluster is older than about $2\,\mathrm{Myr}$,
the most massive star is likely a binary product (note that our 
calculations are for a metallicity of $Z=0.02$ 
while the R136 cluster in the Large Magellanic Cloud has a lower metallicity --- 
so the above numbers will slightly change for the appropriate metallicity 
but are good enough for this estimate).
Because it is not clear whether the most massive star in R136 is a binary product
or not, we explore both possibilities.

With Monte Carlo simulations, we investigate the likelihood of finding 
the observed $280$--$320\,\msun$ stars \citep{2010MNRAS.408..731C} in R136.
We randomly sample R136-like star clusters for different adopted stellar 
upper mass limits $M_{\rm up}$ using the observed IMF 
slope \citep{1998ApJ...493..180M} of $\Gamma=-1.35$, a binary fraction of $70\%$
and that R136 contains about $650$~stellar systems more massive than $10\,\msun$ 
\citep{1997AJ....113.1691H}. We then compute the average number of stars that 
are initially more massive than a given mass $M$, 
$\left\langle N_{\geq M}\right\rangle$, and the probability 
that at least one star is more massive than $M$, $P_{\geq M}$, by repeating 
each experiment $1000$ times (the quoted errors are $1\sigma$ standard deviations). 
The average numbers and probabilities for the case that 
binary interactions did not yet take place are summarised
in Table~\ref{tab:upper-mass-limit-single-stars}. For the case that binary interactions
already took place, we assume that all massive binaries with initial 
periods $P_{\rm i}\leq 5\,{\rm d}$ interact by mass transfer (which happens within $2$--$3\,{\rm Myr}$)
and that the post-interaction mass is $90\%$ of the total binary mass. 
The corresponding average numbers and probabilities for this case 
can be found in Table~\ref{tab:upper-mass-limit-binary-stars}.
In both Tables.~\ref{tab:upper-mass-limit-single-stars} and~\ref{tab:upper-mass-limit-binary-stars},
we also give the results for less massive clusters with $100$ and $350$
stellar systems initially exceeding $10\,\msun$.

Through binary mergers, stars of up to $300\,\msun$ can be produced if 
the star formation process stops at an upper mass of $150\,\msun$.
However, this scenario requires equal mass O-type binaries
which are rare \citep{2012Sci...337..444S,2013A&A...550A.107S}.
We find that with an upper mass of $150\,\msun$, the probability of forming 
stars in excess of $275\,\msun$ in R136 is zero\footnote{given our assumptions, 
the maximum achievable post-interaction mass is $90\%$ of the total system 
mass, i.e. $270\,\msun$ for $M_\mathrm{up}=150\,\msun$}. 
With an upper mass limit of $175\,\msun$, the probability of forming at least 
one star of mass $M\geq275\,\msun$ increases to $7.0\%$, and for an upper mass 
limit of $\sim200\,\msun$, the probability of forming at least one star 
exceeding $275\,\msun$ and $300\,\msun$ is $22.8\%$ and $10.6\%$, respectively.
So, $200\,\msun$ provides a lower limit on the maximum stellar birth mass.

It is also unlikely that the upper mass limit exceeds $350\,\msun$
because then the probability of forming one star above $350\,\msun$ 
by binary mass transfer increases to $50.5\%$
but such massive stars are not observed. We conclude that an upper mass limit 
in the range of about $200$--$350\,\msun$ is needed to explain the most massive stars in R136 by binary evolution. 

Dynamically induced stellar coalescence was proposed as a mechanism to produce the 
very massive stars in R136 \citep{1999A&A...348..117P,2012MNRAS.426.1416B}. 
However, N-body simulations of dynamically induced stellar coalescence typically only produce
one to two stars exceeding $200\,\msun$ for R136 adopting an upper mass limit of $150\,\msun$ \citep{2012MNRAS.426.1416B},  
i.e. fewer than observed in R136. Furthermore, the rate of 
dynamically induced mergers in these simulations should be viewed as an upper limit only. 
Observational results indeed favour a larger half-mass radius \citep{1995ApJ...448..179H,2012A&A...546A..73H}, 
hence a
lower density 
compared to the simulation assumptions. 
Similarly, adopting the recent measurements of the orbital distributions of massive 
binaries \citep{2012Sci...337..444S,2013A&A...550A.107S} further decreases 
the number of possible dynamical mergers that can overcome the $150\,\msun$ limit by a factor of $3.5$ to $4.0$. 
It appears thus unlikely that dynamically induced stellar coalescence is sufficiently efficient to
explain the origin of the very massive stars in R136 if the upper mass limit is $150\,\msun$.

As mentioned above, it is also possible that the four massive stars in R136 
were born with their deduced initial masses and did not gain mass by other means.
This provides then an upper limit on the maximum stellar birth mass.
The most massive star found in R136 has an initial mass of $320_{-40}^{+100}\,\msun$ 
\citep{2010MNRAS.408..731C} --- hence, the upper mass limit has to be
at least of this order, i.e. $\gtrsim 280\,\msun$. 
This initially $320\,\msun$ star allows us to exclude 
an upper mass limit of $M_{{\rm up}}=10^{4}\,\msun$ with $96\%$ confidence 
because we expect to find $3.2\pm1.8$ stars that initially 
exceed $500\,\msun$ in this case --- while no such star is observed. 
However, it becomes more difficult to exclude an upper mass limit of 
$500\,\msun$ or less because the probability of
finding \emph{no} star that initially exceeds $350\,\msun$ ($1-P_{\geq350}$) is about $13\%$; 
in other words, no star would initially exceed $350\,\msun$ in about every tenth R136-like 
star cluster for an upper mass limit of $500\,\msun$. The probability increases 
further to $39\%$ for an upper mass limit of $400\,\msun$. We conclude that stochastic 
sampling effects are important even in the richest massive star clusters in the Local Group. 

Altogether, we find that current data does not
exclude an upper mass limit as high as $400$--$500\,\msun$ if binary interactions are 
neglected. However, the most massive star in R136 is a binary product with a 
probability of $\gtrsim 40$--$60\%$. Including effects of close 
binary evolution, an initial stellar upper mass limit of at least $200\,\msun$
is required to explain the observed stars with apparent initial masses of about $300\,\msun$.
The upper mass limit is thus in the range $200$--$500\,\msun$,
thereby solving the maximum mass problem.

\section{Uncertainties}\label{sec:discussion}
There are several sources of uncertainty that affect theoretical and observed mass functions 
and hence e.g. our cluster ages derived from them. It is important to understand the uncertainties
to estimate their influence on our conclusions and the derived quantities. 
The conclusion that binary effects shape the upper end of the stellar mass function remains unaffected.
In Sec.~\ref{sec:uncertainties-models}, we discuss modelling uncertainties 
because of the fitting procedure, stellar wind mass loss, binary star evolution
and rotation. Observational uncertainties like the influence of different reddening 
laws on derived stellar masses of stars in the Galactic Center are discussed 
in Sec.~\ref{sec:uncertainties-observations}. We discuss the influence of 
dynamical interactions on stellar mass functions in Sec.~\ref{sec:dynamical-effects}.
Star formation histories that are different from single starbursts are considered in
Sec.~\ref{sec:sfh} to investigate whether such scenarios are also consistent with
the observed age spread among the most massive stars and the resulting stellar mass functions.

\subsection{Modelling uncertainties}\label{sec:uncertainties-models}
\subsubsection{Fitting uncertainties}\label{sec:fitting-uncertainties}
Stars in the wind-mass-loss peak of the mass function will very soon
leave the main sequence. The mass of these turn-off stars is a
sensitive function of cluster age especially for massive stars, which radiate close
to the Eddington limit. Massive stars have lifetimes which depend only weakly on
mass and hence a small change in age corresponds to a large change
in mass. We cannot reproduce the observed mass functions of Arches
and Quintuplet if we change the age of our models in Fig.~\ref{fig:observations}
by more than $0.2$--$0.3\,\mathrm{Myr}$. We therefore adopt $0.3\,\mathrm{Myr}$
as the age uncertainty associated with our fitting. 

The initial binary fraction is best constrained by the number
of stars in the mass function tail because it consists only of either
post-interaction or pre-interaction, unresolved binaries. In contrast, the wind-mass-loss
peak changes little with the binary fraction. Our observational sample
is limited by the exclusion of WNh stars in both Arches and Quintuplet
because no relieable masses of the WNh stars could be determined (see Sec.~\ref{sec:observations}.
We thus cannot determine the primordial binary fractions accurately,
especially because the tail of the Quintuplet mass function is not
very pronounced. Increasing the age of our Quintuplet model by $0.1\,\mathrm{Myr}$ allows
for $100\%$ binaries while maintaining a satisfactory, albeit slightly
inferior to the best, fit to the mass function. Both clusters are
thus consistent with having the same primordial binary fraction.

\subsubsection{Stellar wind mass loss}\label{sec:wind-mass-loss}
Our wind mass loss prescription \citep{1990A&A...231..134N} slightly
underestimates stellar winds compared to the latest predictions \citep{2000A&A...362..295V,2001A&A...369..574V}.
Compared to the most recent stellar evolution models of \citet{2012A&A...537A.146E} that use
the prescriptions of \citet{2000A&A...362..295V,2001A&A...369..574V}, we find that our turn-off
masses agree to within $2-3\%$ for initial masses $\lesssim50\,\msun$,
while in more massive stars our turn-off mass is up
to $15\%$ larger, mainly because of the applied Wolf-Rayet wind mass loss 
rates in \citet{2012A&A...537A.146E}. 

The widths of the bins in our model mass functions are $0.04\,\mathrm{dex}$, i.e.\
masses differ by about $10\%$ from bin to bin. The observed mass functions
have bin sizes of $0.2\,\mathrm{dex}$, i.e.\ masses are different by $59\%$ from bin
to bin. Wind mass loss prescriptions that lead to stellar masses at the end of the MS that 
differ by only a few percent result in indistinguishable mass functions ---
our mass functions and conclusions are therefore essentially independent of whether the empirical wind mass
loss prescription of \citet{1990A&A...231..134N} or the theoretical prescription
of \citet{2000A&A...362..295V,2001A&A...369..574V} are used.

Augmenting our wind loss rate by $70\%$, we find
that an initially $85\,\msun$ star has a turn-off mass of about $49\,\msun$
which matches the recent stellar models by \citet{2012A&A...537A.146E}
(compared to $\sim58\,\msun$ in our standard model). 
With the enhanced wind mass loss rate, the Arches
wind-mass-loss peak corresponds to an initially $\sim85\,\msun$ star
with a main-sequence lifetime of $3.3\,\mathrm{Myr}$, compared to
$70\,\msun$ and $3.5\,\mathrm{Myr}$ respectively in our standard
model. The Quintuplet wind-mass-loss peak comes from initially $\sim40\,\msun$
stars for which the uncertainty in wind mass loss is $<3\%$. Our
Quintuplet age estimate is thus robust with respect to the
wind mass loss rate uncertainty.

\subsubsection{Binary star evolution}\label{sec:binary-star-evolution}
Our understanding of binary star evolution in general is subject to uncertainties. Uncertainties
that directly influence the shape of the mass function tails are discussed in 
\citet{Schneider+2013a}. A further, more quantitative discussion of uncertainties in binary star
evolution is found in \citet{2013ApJ...764..166D} and \citet{de-Mink+2013b}.
Here, we restrict ourselves to MS stars, i.e.\ to mergers of two MS stars 
and mass transfer onto MS stars. Mergers that involve a post-MS star 
form a post-MS object and are thus not considered here.

We assume that two MS stars merge if the mass ratio of the accretor to donor star
is less than $0.56$. This threshold is calibrated against the detailed binary
models of \citet{2007A&A...467.1181D} and is of limited relevance
to our results: if a binary does not merge but instead transfers mass (or vice versa),
the accretor becomes massive because the mass transfer efficiency of MS stars
is high \citep[e.g.][]{2001A&A...369..939W,2012ARA&A..50..107L}. In either case, 
the mass gainer will be a massive star \citep{2013ApJ...764..166D}. The expected binary fraction 
of stars in the tail of the mass functions however changes: a lower
critical mass ratio leads to fewer MS mergers and hence to a higher binary fraction and vice versa.

The amount of rejuvenation of MS mergers is determined by the amount of mixing of fresh fuel into
the core of the merger product and determines by how much the lifetime of the merger product
is prolonged. The more rejuvenation, the longer the remaining MS lifetime and the more mergers
are expected to be found. We assume that a fraction of $10\%$ of the envelope is mixed 
into the core, resulting into fairly short remaining MS lifetimes of the merger products 
compared to the assumption of complete mixing used in the original \citet{2002MNRAS.329..897H} code.
Recent simulations of massive mergers seem to support the mild mixing as used in 
our work \citep[][and references therein]{2013MNRAS.434.3497G}.

The mass transfer efficiency is important for our results. The more of the transferred 
mass is accreted during RLOF, the larger the final mass of the accreting star. 
The maximum reachable mass of any accretor is given by the total mass of the binary 
(i.e. at most twice the mass of the donor star) and the larger the overall mass transfer efficiency,
the more binary products exceed the most massive
genuine single star. In our models, we limit the mass accretion rate to the thermal
timescale of the accretor, which results in higher mass transfer efficiencies 
the larger the mass ratio and the closer the binary \citep[see][]{Schneider+2013a}.
This idea is motivated by detailed binary evolution models 
\citep[e.g.][]{1976ApJ...206..509U,1977A&A....54..539K,1977PASJ...29..249N,1994A&A...288..475P,2001A&A...369..939W}.

The initial distributions of primary masses, mass ratios and orbital separations 
determine the relative fraction of stars that will merge, transfer mass etc. It turns 
out that the distribution of orbital separations influences the incidence of binary 
products most \citep{2013ApJ...764..166D,de-Mink+2013b} because initially close binaries 
transfer mass on average more efficiently than wider binaries. A distribution 
of initial orbital separations that favours close binaries therefore leads to 
on average more massive binary products than distribution functions that favour initially 
wider binaries.

A more quantitative assessment of the above quoted uncertainties in binary evolution and
initial binary distribution functions reveals that a population of MS stars with 
luminosities $L>10^4\,\lsun$ (i.e. O- and B-stars)
contains about $30^{+10}_{-15}\%$ binary products if continuous star formation is
assumed \citep{de-Mink+2013b}. All in all, depending on the exact assumptions regarding binary evolution, there will
be more or fewer stars in the tail of the mass function. However, the peak-tail structure 
never disappears unless it is assumed that neither MS mergers nor RLOF are able to 
increase stellar masses which is unphysical.

\subsubsection{Stellar rotation}\label{sec:stellar-rotation}
Mixing induced by stellar rotation increases the fuel available
to a star and increases its lifetime. The amount of mixing grows with
increasing mass, increasing rotation rate and decreasing metallicity and may contribute
to the observed age spreads and mass function tails in Arches and Quintuplet. The models
of \citet{2011A&A...530A.115B} show that the MS lifetime of
a $60\,\msun$ star lengthens by $0.2\,\mathrm{Myr}$ and $0.6\,\mathrm{Myr}$
for initial rotational velocities of $300$ and $500\,\mathrm{km}\,\mathrm{s^{-1}}$,
respectively. Assuming that the present day distribution of rotational velocities of Galactic 
O- and B-stars approximately represents the initial distribution, 
no more than $10\%$ and less than $1\%$ of stars would have initial rotation rates exceeding $300$ and $500\,\mathrm{km}\,\mathrm{s}^{-1}$,
respectively, and are thus expected to be influenced significantly by rotational mixing 
\citep[see Table~2 in][and references therein]{2013ApJ...764..166D}. This is small compared to the $40\%$ of all O-stars
that undergo strong binary interaction during their main-sequence evolution
\citep{2012Sci...337..444S}. The present day distribution of rotational velocities is probably
altered e.g. by binary star evolution such that some of the fast rotators
are expected to have gained their fast rotation by binary interactions (\citealt{2013ApJ...764..166D}, but see also \citealt{2013arXiv1309.2929R}).
In this respect, the expected fraction of genuine Galactic single stars that are significantly affected by 
rotational mixing is even smaller than the above quoted fractions.
The effect of rotation is thus only of limited relevance to our results compared
to binary star interactions.

\subsection{Observational uncertainties}\label{sec:uncertainties-observations}
There are two steps involved in determining stellar masses from photometric
observations that contribute to the uncertainties of the derived individual stellar masses. 
The first step involves the conversion of the observed apparent magnitudes (fluxes) to 
absolute magnitudes and luminosities, respectively, taking --- amongst others --- 
the distance and extinction into account. 
The second step involves the conversion of luminosities to stellar masses. 
This step relies upon mass-luminosity relations that depend on 
(in general a priori unknown) stellar ages and the applied stellar models. 
In this section, we estimate the uncertainty on individual stellar masses introduced
by these two steps for stars in Arches and Quintuplet. Once we know the uncertainties,
we can apply them to the turn-off masses derived from the wind mass-loss peak in the mass function 
to find the corresponding uncertainty in cluster age.

\begin{figure}
\center
\includegraphics[width=0.46\textwidth]{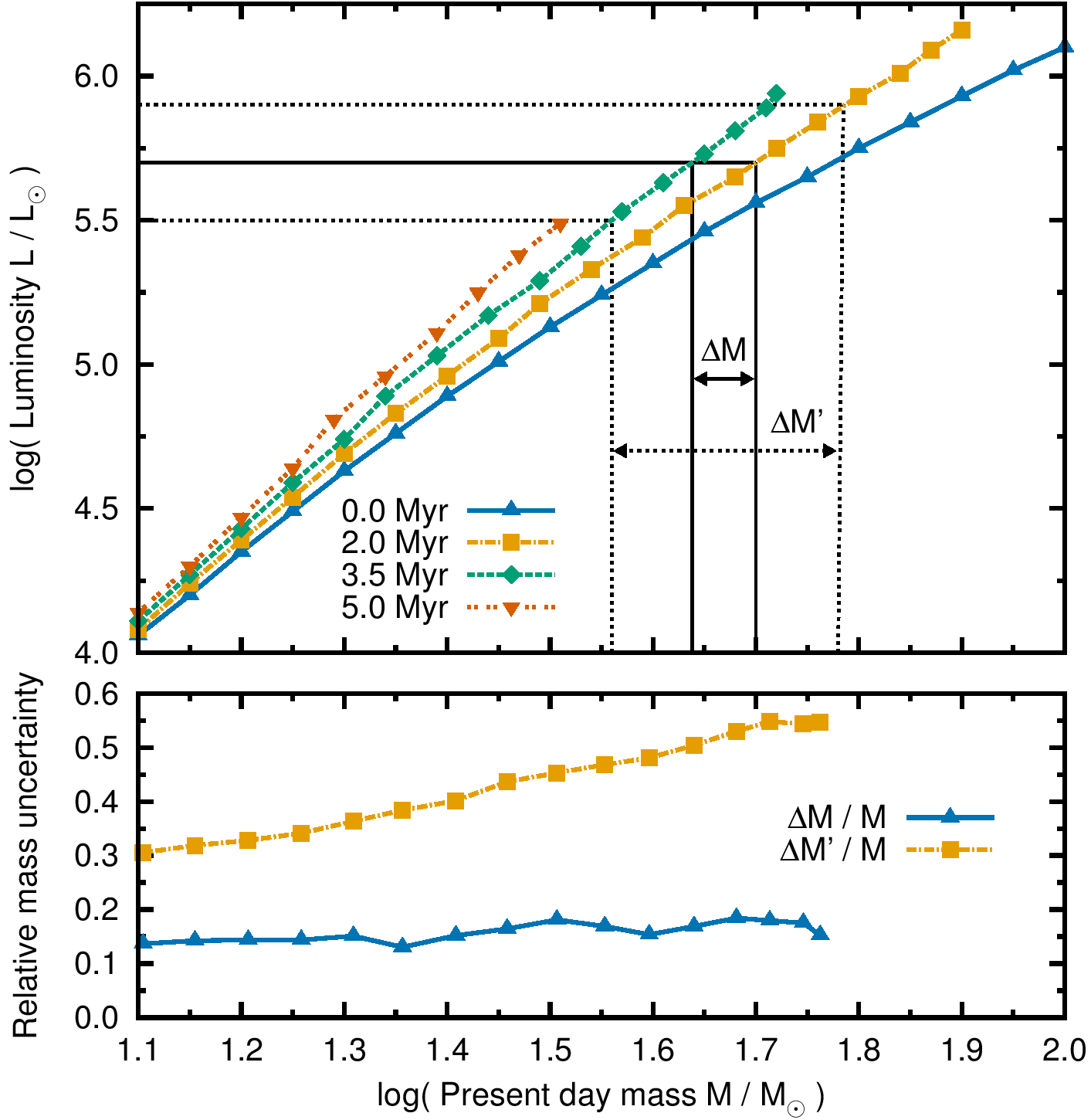}
\caption{Uncertainties on derived individual stellar masses introduced by unknown a priori stellar
ages, $\Delta M/M$, and by the combined effect of unknown stellar ages and uncertain luminosities,
$\Delta M'/M$. The upper panel illustrates how we estimate these uncertainties for a star
with luminosity $\log L/\lsun = 5.7\pm0.2$ using the main-sequence mass-luminosity relations of different ages
from our code \citep{2000MNRAS.315..543H}. The lower panel shows the uncertainties as a function of mass.}
\label{fig:mass-uncertainty}
\end{figure}

In the upper panel of Fig.~\ref{fig:mass-uncertainty} we show main-sequence mass-luminosity relations
of Milky Way stars of different ages as used in our code \citep[based on][]{2000MNRAS.315..543H}. Not knowing
the exact age of a star, but only a probable age range (here $2.0\text{--}3.5\,\mathrm{Myr}$),
introduces an uncertainty, $\Delta M$, on the derived individual stellar masses (cf. lower panel in 
Fig.~\ref{fig:mass-uncertainty}). The more evolved a star and the more massive, the
larger is the uncertainty. If we additionally include the uncertainty in the 
luminosity (here $\pm0.2\,\mathrm{dex}$; see below) the uncertainty in the stellar mass grows to $\Delta M'$.
The uncertainty in luminosity is the dominant contribution here.

The uncertainties presented in Fig.~\ref{fig:mass-uncertainty} are tailored to stars in the
Arches cluster where stellar masses have been derived by \citet{2005ApJ...628L.113S} using a 
$2\,\mathrm{Myr}$ isochrone (mass-luminosity relation) whereas our analysis reveals an age of about 
$3.5\,\mathrm{Myr}$ --- hence the $2.0\text{--}3.5\,\mathrm{Myr}$ age range for the mass-luminosity relations.
Deriving individual stellar masses using two different extinction laws, \citet{2013A&A...556A..26H} 
find that stellar masses can deviate by up to $30\%$ when either a \citet{2009ApJ...696.1407N} 
extinction law or the traditional \citet{1985ApJ...288..618R} law towards the Galactic Center are used. 
As the appropriate extinction law towards the Galactic Center line of sight is still a matter of debate, 
we adopt uncertainties of $\pm0.2\,\mathrm{dex}$ on luminosities to be in line with the work of \citet{2013A&A...556A..26H}.
The situation is similar in Quintuplet: we find slightly smaller uncertainties on 
derived stellar masses for uncertainties of $\pm0.2\,\mathrm{dex}$ on luminosities 
and an age range of $4\text{--}5\,\mathrm{Myr}$ than given in Fig.~\ref{fig:mass-uncertainty}. 
To be conservative, we adopt that individual stellar masses are uncertain by 
$\pm30\%$ (i.e. $\Delta M'/M=60\%$) in both clusters, which agrees with
the diversity of mass estimates in the literature for stars in Arches.

\citet{2013A&A...556A..26H} find stellar masses that are up to $30\%$ less than those of our analysis 
for a \citet{2009ApJ...696.1407N} extinction law.
An independent study of the Arches present-day mass function \citep{2006ApJ...653L.113K}
shows a similar peak-tail structure at the high mass end using a different
extinction law and the derived masses are comparable to those of our
analysis \citep{2005ApJ...628L.113S}. Masses in excess of $150\,\msun$ --- i.e. larger than our
masses --- have been suggested for the most luminous stars in the Arches cluster by
\citet{2010MNRAS.408..731C}. Hence, an uncertainty of $\pm30\%$ applied
to our adopted stellar masses 
covers the complete range of suggested masses for stars in Arches.

The overall structure of the mass functions of Arches and Quintuplet --- 
a stellar-wind peak and binary tail --- is robust to the mentioned 
uncertainties. We also find the peak-tail structure in the mass function 
of \citet{2013A&A...556A..26H} and in mass functions constructed from the photometric 
data of \citet{2005ApJ...628L.113S} and \citet{2012A&A...540A..57H}, respectively, 
using isochrones of different ages. The whole mass functions shift in mass, but the
relative structure stays the same. The reason is that different extinction laws
or isochrones systematically influence all stars in a similar way and do not introduce 
differential effects.

Applying $\pm30\%$ uncertainties on stellar masses leads to turn-off 
masses of the Arches and Quintuplet of about 
$70\pm21\,\msun$ and $40\pm12\,\msun$, with associated age uncertainties of 
$\pm0.6$ and $\pm1.1\,\mathrm{Myr}$ respectively.

\subsection{Dynamical interactions in star clusters}\label{sec:dynamical-effects}
The observed present-day mass functions are influenced 
by dynamical cluster evolution \citep[][and references therein]{2010MNRAS.409..628H}.  
Flat mass function slopes of the order of $\Gamma = -0.7$, 
compared to a Salpeter IMF slope of $\Gamma = -1.35$, are likely a consequence of mass 
segregation in which massive stars sink towards the cluster center 
For the Arches cluster, \citet{2013A&A...556A..26H}
show that a dynamical model with a standard Salpeter IMF 
explains the steepening of the IMF slope towards larger
distances from the cluster center. While the conclusion is more elusive for the dispersed
Quintuplet population, the similarity of the mass function slopes 
in the inner regions of both clusters and the older age of Quintuplet suggest that
similar processes have shaped the present-day mass function
of this cluster as well. 

We further investigate whether dynamical interactions in star clusters
can also cause the peak in the observed mass functions.
Mass segregation flattens the high-mass end of the mass function without producing a
peak \citep{2007MNRAS.378L..29P} while dynamical ejection of stars works
on all stars with an ejection efficiency monotonically increasing with
mass \citep{2011Sci...334.1380F,2012ApJ...751..133P,2012ApJ...746...15B}
as confirmed by the observed Galactic fraction of runaway O-stars which
is larger than that of runaway B-stars \citep{1986ApJS...61..419G,1991AJ....102..333S}.
Such a smoothly increasing ejection efficiency does not create
a peak in the mass function but can give rise to a tail which is however
not seen in the mass functions of N-body simulations of the Arches 
cluster \citep{2007MNRAS.378L..29P,2010MNRAS.409..628H}.

\subsection{Star formation histories}\label{sec:sfh}
Star formation is not an instantaneous process but lasts a finite amount of time. 
Observations show that there is an empirical
relationship between the duration of star formation and the crossing
time of star clusters \citep{2000ApJ...530..277E}. For example, in Westerlund~1 and NGC~3603YC the age difference
among stars less massive than about $11.5\,\msun$ and $6.5\,\msun$ shows that they were formed
within at most $0.4$ and $0.1\,\mathrm{Myr}$ \citep{2012ApJ...750L..44K}, respectively, which
compare well to the cluster crossing times of $0.3\,{\rm Myr}$ \citep{2008A&A...478..137B}
and $0.1\,\mathrm{Myr}$ \citep{2010IAUS..266...24P}. 

The core of the Arches cluster has a radius of about $0.23\,{\rm pc}$ \citep{2002ApJ...581..258F}
and a velocity dispersion of $5.7\,{\rm km}\,{\rm s}^{-1}$ \citep{2012ApJ...751..132C}
at a distance of $8.0\,{\rm kpc}$ corresponding
to a crossing time, and hence a duration of star formation, of about
$0.04\,{\rm Myr}$. A similar estimate for the Quintuplet cluster is
more uncertain because the core radius and especially the velocity dispersion 
are not well known. If we assume that the observed central region of Quintuplet with a radius of
about $0.5\,{\rm pc}$ \citep{2012A&A...540A..57H} corresponds to the core radius 
and that the velocity dispersion is about $17\,{\rm km}\,{\rm s}^{-1}$ \citep{2009A&A...494.1137L},
the crossing time is $0.03\,{\rm Myr}$. The mass functions produced by such short periods
of star formation are indistinguishable from an instantaneous starburst.
The apparent age spread among the most
luminous stars in Arches and Quintuplet is about $1$--$1.5\,{\rm Myr}$ \citep{2008A&A...478..219M,2012A&A...540A..14L}
and hence much larger than the estimated star formation periods.

In Appendix~\ref{sec:sfh-cont} we investigate whether a star formation history that is 
different from a single starburst can also explain the observed peak-tail structure in the mass functions
of Arches and Quintuplet. We find that this is possible with e.g. a two-stage starburst but 
the age spread among the most massive stars would then be inconsistent with observations.

\section{Conclusions}\label{sec:conclusions}
Massive stars rapidly change their mass, thereby altering the stellar mass function.
Stellar wind mass loss reduces stellar masses such that stars accumulate 
near the high mass end of present day mass functions, creating a bump whose
position reveals the mass of the turn-off stars and hence the age of young star clusters \citep{Schneider+2013a}.
Binary stars are frequent and important for massive star evolution because
of binary mass exchange \citep{2012Sci...337..444S}: mass transfer 
and stellar mergers increase stellar masses and create a
tail of rejuvenated binary products at the high mass end of mass functions \citep{Schneider+2013a}.
We model the observed mass functions of the young Arches and 
Quintuplet star clusters \citep{2005ApJ...628L.113S,2012A&A...540A..57H}
using a rapid binary evolution code \citep{2002MNRAS.329..897H,2004MNRAS.350..407I,2006A&A...460..565I,2009A&A...508.1359I} 
to identify these two features and to address two pressing controvercies:
\begin{enumerate}
	\item \emph{The cluster age problem:} the most massive stars in Arches and Quintuplet, the WNh stars, 
	appear to be younger than the less massive O-stars \citep{2008A&A...478..219M,2012A&A...540A..14L}. 
	This is not expected from star cluster formation \citep{2000ApJ...530..277E,2012ApJ...750L..44K} 
	but well known in older clusters under the blue straggler phenomenon.
	\item \emph{The maximum mass problem:} a stellar upper mass limit of $150\,\msun$ is observationally determined 
	\citep{2004MNRAS.348..187W,2005Natur.434..192F,2005ApJ...620L..43O,2006MNRAS.365..590K}. In contrast,
	the supernova SN~2007bi is thought to be a pair-instability supernova from an initially
	$250\,\msun$ star \citep{2009Natur.462..624G,2009Natur.462..579L} and four stars greatly exceeding this limit are 
	found in the R136 cluster in the Large Magellanic Cloud \citep{2010MNRAS.408..731C}.
\end{enumerate}
We identify the peak and tail in the observed mass functions of the Arches and Quintuplet
clusters. By fitting our models to the observations, we determine the ages
of Arches and Quintuplet to be $3.5\pm0.7\,\mathrm{Myr}$ and $4.8\pm1.1\,\mathrm{Myr}$,
respectively. This solves the age problem because the most massive stars in
Arches and Quintuplet are rejuvenated products of binary mass transfer
and the ages derived from these stars are therefore significantly underestimated.
While our age error bars are still large --- mostly because of uncertain absolute magnitudes ---
our method removes the ambiguity in the age determination. For the age determination
of older star clusters, blue stragglers are eminently disregarded when 
isochrones are fitted to the turn-off in Hertzsprung-Russell diagrams. Our analysis shows
that for young star clusters, where the higher fraction of interacting binaries 
produces even more blue stragglers, they obviously need to be disregarded as well in
order to derive the correct cluster age.

Even without modelling the observed mass functions, the ages and also the
mass lost by the turn-off stars during their main-sequence evolution
can be determined from the mass function alone. Refilling the mass function above
the present day mass of the turn-off stars with the number of excess stars 
in the wind-mass-loss peak gives the initial mass of the turn-off stars 
and hence the cluster age. The derived ages agree with the more accurate ages
from detailed modelling of the observed mass functions. According to this new method, 
the turn-off stars in Arches lost $12$--$17\,\msun$ of their initial $62$--$72\,\msun$ 
and $4$--$8\,\msun$ of their initial $36$--$47\,\msun$ in Quintuplet.
This is the first direct measurement of stellar wind mass loss which does
not rely on derivations of stellar wind mass loss rates.

Monte Carlo experiments to investigate the effects of stochastic sampling
show that the most massive star in the Arches and Quintuplet clusters is expected to be
a rejuvenated product of binary mass transfer after $1.0\pm0.7\,\mathrm{Myr}$
and $1.7\pm1.0\,\mathrm{Myr}$, respectively. 
At their present age, the probability that the most massive star in
Arches and Quintuplet is a product of binary mass exchange is $>99.9\%$
and the most massive $9.2\pm3.0$ and $7.5\pm2.8$ stars in Arches and Quintuplet, respectively, 
are expected to be such rejuvenated binary products.

Our findings have implications for the maximum mass problem. The Arches
cluster is older than previously thought and its most massive stars
are most likely binary products. The mass function is thus truncated
by finite stellar lifetimes and not by an upper mass limit.
To constrain a potential stellar upper mass limit, we consider the massive cluster R136
in the Large Magellanic Cloud which is thought to be so young that its
initially most massive stars are still alive today.
We find that the most massive star is a binary product with a probability of $>40\%$, 
depending on the exact, albeit yet uncertain cluster age (Sec.~\ref{sec:upper-mass-limit}). Assuming 
binaries already interacted, a stellar upper mass limit of at 
least $200\,\msun$ is needed to form the observed
$165$--$320\,\msun$ stars in R136. It can also not exceed $350\,\msun$ because then
the probability of forming stars above e.g. $350\,\msun$ becomes larger than about 
$50\%$ --- but such stars are not observed. Assuming that no binary interactions 
changed the masses of the very massive 
stars in R136, a stellar upper mass limit of up to $400$--$500\,\msun$ 
cannot be fully excluded because of stochastic sampling even in this rich star cluster. 
The upper mass limit is thus likely in the range $200$--$500\,\msun$, 
thereby solving the maximum mass problem.

We conclude that the most massive stars in the Universe may be the 
rejuvenated products of binary mass transfer. 
Because of their extreme mass and luminosity, radiation feedback from these stars 
is crucial to observable properties of young stellar
populations, to the state of the interstellar medium around young
stellar clusters and even to the reionization of the Universe after the Big Bang.
Our results have strong implications for understanding star-forming regions nearby and at high redshift 
as observationally derived fundamental properties like initial mass functions
are based on the assumption that the brightest stars are single and less massive than $150\,\msun$. 
These very massive stars are thought to die as pair-instability supernovae 
and produce huge, so far unaccounted contributions to the chemical enrichment of 
nearby and distant galaxies \citep{2012ARA&A..50..107L} and their final explosions 
may be observable throughout the Universe. Understanding the most massive stars 
in young nearby star clusters is an essential step towards investigating these 
exciting phenomena which shape our cosmos.

\begin{acknowledgements}
We thank the referee, Dany Vanbeveren, for carefully reading our manuscript and constructive suggestions.
F.R.N.S. acknowledges the fellowships awarded by the German National Academic Foundation (Studienstiftung)
and the Bonn-Cologne Graduate School of Physics and Astronomy.
R.G.I. would like to thank the Alexander von Humboldt foundation.
S.d.M. acknowledges support by NASA through Hubble Fellowship grant 
HST-HF-51270.01-A awarded by the Space Telescope Science Institute, 
which is operated by the Association of Universities for Research in Astronomy, 
Inc., for NASA, under contract NAS 5-26555 and the Einstein Fellowship program 
through grant PF3-140105  awarded by the Chandra X-ray Center, which is 
operated by the Smithsonian Astrophysical Observatory for NASA under the contract NAS8-03060.
B.H. and A.S. acknowledge funding from the German science foundation 
(DFG) Emmy Noether program under grant STO 496-3/1.
\end{acknowledgements}

\appendix

\begin{table*}[b]
\caption{Results of our Monte Carlo simulations to determine the stellar
upper mass limit without binary interactions.}
\label{tab:upper-mass-limit-single-stars}
\centering
\begin{tabular}{cccccccc}
\tableline 
 &  & \multicolumn{2}{c}{$N_{10}=100$} & \multicolumn{2}{c}{$N_{10}=350$} & \multicolumn{2}{c}{$N_{10}=650$}\tabularnewline
 &  & \multicolumn{2}{c}{$M_\mathrm{cl}\approx 2\times10^4\,\msun$} & \multicolumn{2}{c}{$M_\mathrm{cl}\approx 7\times10^4\,\msun$} & \multicolumn{2}{c}{$M_\mathrm{cl}\approx10^5\,\msun$}\tabularnewline
\noalign{\vskip\doublerulesep}
$M_{{\rm up}}/\msun$ & $M/\msun$ & $\left\langle N_{\geq M}\right\rangle $ & $P_{\geq M}$ & $\left\langle N_{\geq M}\right\rangle $ & $P_{\geq M}$ & $\left\langle N_{\geq M}\right\rangle $ & $P_{\geq M}$\tabularnewline
\tableline
\tableline
$10000$ & $150$ & $2.6\pm1.6$ & $92.0\%$ & $9.0\pm3.0$ & $99.9\%$ & $16.9\pm4.1$ & $>99.9\%$\tabularnewline
 & $200$ & $1.7\pm1.3$ & $81.8\%$ & $6.1\pm2.5$ & $99.8\%$ & $11.3\pm3.3$ & $>99.9\%$\tabularnewline
 & $250$ & $1.3\pm1.1$ & $71.5\%$ & $4.5\pm2.1$ & $98.2\%$ & $8.3\pm2.9$ & $>99.9\%$\tabularnewline
 & $300$ & $1.0\pm1.0$ & $63.4\%$ & $3.4\pm1.8$ & $96.8\%$ & $6.5\pm2.6$ & $99.9\%$\tabularnewline
 & $350$ & $0.8\pm0.9$ & $55.1\%$ & $2.8\pm1.6$ & $94.1\%$ & $5.2\pm2.3$ & $99.6\%$\tabularnewline
 & $400$ & $0.7\pm0.8$ & $48.5\%$ & $2.3\pm1.5$ & $90.6\%$ & $4.4\pm2.1$ & $99.2\%$\tabularnewline
 & $450$ & $0.6\pm0.7$ & $43.0\%$ & $2.0\pm1.4$ & $86.1\%$ & $3.7\pm1.9$ & $98.1\%$\tabularnewline
 & $500$ & $0.5\pm0.7$ & $38.5\%$ & $1.7\pm1.3$ & $81.4\%$ & $3.2\pm1.8$ & $96.3\%$\tabularnewline
\tableline
$1000$ & $150$ & $2.5\pm1.5$ & $92.5\%$ & $8.3\pm2.8$ & $>99.9\%$ & $15.4\pm3.9$ & $>99.9\%$\tabularnewline
 & $200$ & $1.6\pm1.2$ & $81.3\%$ & $5.4\pm2.3$ & $99.5\%$ & $10.0\pm3.2$ & $>99.9\%$\tabularnewline
 & $250$ & $1.1\pm1.0$ & $69.1\%$ & $3.8\pm2.0$ & $97.5\%$ & $7.0\pm2.7$ & $99.9\%$\tabularnewline
 & $300$ & $0.8\pm0.9$ & $58.0\%$ & $2.8\pm1.7$ & $93.2\%$ & $5.2\pm2.3$ & $99.3\%$\tabularnewline
 & $350$ & $0.6\pm0.7$ & $48.4\%$ & $2.1\pm1.4$ & $87.8\%$ & $4.0\pm2.1$ & $97.3\%$\tabularnewline
 & $400$ & $0.5\pm0.7$ & $40.0\%$ & $1.6\pm1.2$ & $80.3\%$ & $3.1\pm1.8$ & $94.4\%$\tabularnewline
 & $450$ & $0.4\pm0.6$ & $32.8\%$ & $1.3\pm1.1$ & $72.6\%$ & $2.5\pm1.6$ & $91.2\%$\tabularnewline
 & $500$ & $0.3\pm0.5$ & $27.4\%$ & $1.0\pm1.0$ & $65.0\%$ & $2.0\pm1.4$ & $86.4\%$\tabularnewline
\tableline
$500$ & $150$ & $2.2\pm1.5$ & $88.5\%$ & $7.3\pm2.6$ & $>99.9\%$ & $13.3\pm3.5$ & $>99.9\%$\tabularnewline
 & $200$ & $1.3\pm1.2$ & $73.0\%$ & $4.4\pm2.1$ & $98.5\%$ & $8.0\pm2.7$ & $>99.9\%$\tabularnewline
 & $250$ & $0.9\pm0.9$ & $57.4\%$ & $2.8\pm1.7$ & $94.5\%$ & $5.1\pm2.2$ & $99.5\%$\tabularnewline
 & $300$ & $0.5\pm0.7$ & $42.5\%$ & $1.9\pm1.3$ & $84.8\%$ & $3.3\pm1.8$ & $96.3\%$\tabularnewline
 & $350$ & $0.3\pm0.6$ & $28.7\%$ & $1.1\pm1.0$ & $68.1\%$ & $2.0\pm1.4$ & $87.3\%$\tabularnewline
 & $400$ & $0.2\pm0.4$ & $18.0\%$ & $0.7\pm0.8$ & $48.9\%$ & $1.1\pm1.0$ & $68.7\%$\tabularnewline
 & $450$ & $0.1\pm0.3$ & $7.9\%$ & $0.3\pm0.5$ & $24.6\%$ & $0.5\pm0.7$ & $39.0\%$\tabularnewline
 & $500$ & $0.0\pm0.0$ & $0.0\%$ & $0.0\pm0.0$ & $0.0\%$ & $0.0\pm0.0$ & $0.0\%$\tabularnewline
\tableline
$400$ & $150$ & $1.9\pm1.4$ & $84.2\%$ & $6.7\pm2.5$ & $99.9\%$ & $12.6\pm3.5$ & $>99.9\%$\tabularnewline
 & $200$ & $1.0\pm1.0$ & $64.9\%$ & $3.7\pm1.9$ & $97.8\%$ & $7.1\pm2.7$ & $>99.9\%$\tabularnewline
 & $250$ & $0.6\pm0.8$ & $43.1\%$ & $2.2\pm1.4$ & $88.6\%$ & $4.1\pm2.1$ & $97.5\%$\tabularnewline
 & $300$ & $0.3\pm0.6$ & $26.5\%$ & $1.1\pm1.0$ & $69.2\%$ & $2.2\pm1.5$ & $89.9\%$\tabularnewline
 & $350$ & $0.1\pm0.3$ & $11.4\%$ & $0.5\pm0.7$ & $36.3\%$ & $0.9\pm1.0$ & $60.6\%$\tabularnewline
 & $400$ & $0.0\pm0.0$ & $0.0\%$ & $0.0\pm0.0$ & $0.0\%$ & $0.0\pm0.0$ & $0.0\%$\tabularnewline
\tableline
$300$ & $150$ & $1.7\pm1.3$ & $81.5\%$ & $5.6\pm2.3$ & $99.7\%$ & $10.4\pm3.2$ & $>99.9\%$\tabularnewline
 & $200$ & $0.8\pm0.9$ & $53.8\%$ & $2.6\pm1.6$ & $93.1\%$ & $4.9\pm2.2$ & $99.1\%$\tabularnewline
 & $250$ & $0.3\pm0.6$ & $26.2\%$ & $1.0\pm1.0$ & $61.4\%$ & $1.8\pm1.3$ & $84.9\%$\tabularnewline
 & $300$ & $0.0\pm0.0$ & $0.0\%$ & $0.0\pm0.0$ & $0.0\%$ & $0.0\pm0.0$ & $0.0\%$\tabularnewline
\tableline
$200$ & $150$ & $0.9\pm0.9$ & $56.7\%$ & $3.0\pm1.7$ & $95.9\%$ & $5.7\pm2.4$ & $99.8\%$\tabularnewline
 & $200$ & $0.0\pm0.0$ & $0.0\%$ & $0.0\pm0.0$ & $0.0\%$ & $0.0\pm0.0$ & $0.0\%$\tabularnewline
\tableline
\end{tabular}
\tablecomments{Given are the average number of stars $\left\langle N_{\geq M}\right\rangle$ 
initially more massive than $M$ and the probability $P_{\geq M}$ that at least 
one star is initially more massive than $M$ for stochastically sampled star 
clusters as a function of the stellar upper mass limit, $M_{{\rm up}}$, and the 
number of stars, $N_{10}$, more massive than $10\,\msun$. The corresponding total 
cluster masses, $M_\mathrm{cl}$, extrapolated with a Kroupa IMF \citep{2001MNRAS.322..231K} 
down to $0.08\,\msun$ are also provided. All stars are assumed to 
be effectively single, i.e. that no binary interactions took place.}
\end{table*}

\begin{table*}[H]
\caption{Results of our Monte Carlo simulations to determine the stellar
upper mass limit with binary interactions.}
\label{tab:upper-mass-limit-binary-stars}
\centering
\begin{tabular}{cccccccc}
\tableline 
 &  & \multicolumn{2}{c}{$N_{10}=100$} & \multicolumn{2}{c}{$N_{10}=350$} & \multicolumn{2}{c}{$N_{10}=650$}\tabularnewline
 &  & \multicolumn{2}{c}{$M_\mathrm{cl}\approx 2\times10^4\,\msun$} & \multicolumn{2}{c}{$M_\mathrm{cl}\approx 7\times10^4\,\msun$} & \multicolumn{2}{c}{$M_\mathrm{cl}\approx10^5\,\msun$}\tabularnewline
\noalign{\vskip\doublerulesep}
$M_{{\rm up}}/\msun$ & $M/\msun$ & $\left\langle N_{\geq M}\right\rangle $ & $P_{\geq M}$ & $\left\langle N_{\geq M}\right\rangle $ & $P_{\geq M}$ & $\left\langle N_{\geq M}\right\rangle $ & $P_{\geq M}$\tabularnewline
\tableline
\tableline
$400$ & $150$ & $2.2\pm1.5$ & $87.7\%$ & $7.6\pm2.8$ & $99.9\%$ & $14.2\pm3.6$ & $>99.9\%$\tabularnewline
 & $200$ & $1.3\pm1.1$ & $70.9\%$ & $4.4\pm2.2$ & $98.4\%$ & $8.1\pm2.8$ & $>99.9\%$\tabularnewline
 & $250$ & $0.8\pm0.9$ & $51.4\%$ & $2.6\pm1.6$ & $92.9\%$ & $4.8\pm2.2$ & $99.3\%$\tabularnewline
 & $300$ & $0.5\pm0.7$ & $36.1\%$ & $1.5\pm1.2$ & $79.1\%$ & $2.8\pm1.7$ & $93.3\%$\tabularnewline
 & $350$ & $0.2\pm0.5$ & $20.8\%$ & $0.8\pm0.9$ & $56.2\%$ & $1.5\pm1.2$ & $75.2\%$\tabularnewline
 & $400$ & $0.1\pm0.3$ & $7.1\%$ & $0.3\pm0.5$ & $24.6\%$ & $0.5\pm0.7$ & $39.8\%$\tabularnewline
 & $450$ & $0.0\pm0.2$ & $4.7\%$ & $0.2\pm0.5$ & $18.5\%$ & $0.3\pm0.6$ & $26.2\%$\tabularnewline
 & $500$ & $0.0\pm0.2$ & $2.7\%$ & $0.1\pm0.3$ & $11.5\%$ & $0.2\pm0.4$ & $16.3\%$\tabularnewline
 & $550$ & $0.0\pm0.1$ & $1.6\%$ & $0.1\pm0.3$ & $6.2\%$ & $0.1\pm0.3$ & $9.4\%$\tabularnewline
 & $600$ & $0.0\pm0.1$ & $0.7\%$ & $0.0\pm0.2$ & $2.9\%$ & $0.0\pm0.2$ & $4.3\%$\tabularnewline
 & $650$ & $0.0\pm0.1$ & $0.3\%$ & $0.0\pm0.1$ & $0.8\%$ & $0.0\pm0.1$ & $1.5\%$\tabularnewline
 & $700$ & $0.0\pm0.0$ & $0.1\%$ & $0.0\pm0.0$ & $0.2\%$ & $0.0\pm0.0$ & $0.1\%$\tabularnewline
 & $750$ & $0.0\pm0.0$ & $0.0\%$ & $0.0\pm0.0$ & $0.0\%$ & $0.0\pm0.0$ & $0.0\%$\tabularnewline
 & $800$ & $0.0\pm0.0$ & $0.0\%$ & $0.0\pm0.0$ & $0.0\%$ & $0.0\pm0.0$ & $0.0\%$\tabularnewline
\tableline
$350$ & $150$ & $2.1\pm1.4$ & $87.0\%$ & $7.2\pm2.6$ & $>99.9\%$ & $13.6\pm3.6$ & $>99.9\%$\tabularnewline
 & $200$ & $1.1\pm1.1$ & $68.4\%$ & $3.9\pm1.9$ & $98.2\%$ & $7.6\pm2.6$ & $99.9\%$\tabularnewline
 & $250$ & $0.6\pm0.8$ & $46.9\%$ & $2.1\pm1.4$ & $88.2\%$ & $4.2\pm2.0$ & $98.8\%$\tabularnewline
 & $300$ & $0.3\pm0.5$ & $24.4\%$ & $1.1\pm1.0$ & $66.1\%$ & $2.1\pm1.5$ & $88.0\%$\tabularnewline
 & $350$ & $0.1\pm0.3$ & $8.3\%$ & $0.3\pm0.6$ & $28.2\%$ & $0.7\pm0.8$ & $50.5\%$\tabularnewline
 & $400$ & $0.0\pm0.2$ & $4.2\%$ & $0.2\pm0.4$ & $16.4\%$ & $0.4\pm0.6$ & $33.0\%$\tabularnewline
 & $450$ & $0.0\pm0.1$ & $2.1\%$ & $0.1\pm0.3$ & $8.3\%$ & $0.2\pm0.5$ & $20.7\%$\tabularnewline
 & $500$ & $0.0\pm0.1$ & $1.1\%$ & $0.0\pm0.2$ & $3.2\%$ & $0.1\pm0.3$ & $10.4\%$\tabularnewline
 & $550$ & $0.0\pm0.1$ & $0.3\%$ & $0.0\pm0.1$ & $1.6\%$ & $0.0\pm0.2$ & $2.9\%$\tabularnewline
 & $600$ & $0.0\pm0.0$ & $0.0\%$ & $0.0\pm0.0$ & $0.0\%$ & $0.0\pm0.0$ & $0.0\%$\tabularnewline
 & $650$ & $0.0\pm0.0$ & $0.0\%$ & $0.0\pm0.0$ & $0.0\%$ & $0.0\pm0.0$ & $0.0\%$\tabularnewline
 & $700$ & $0.0\pm0.0$ & $0.0\%$ & $0.0\pm0.0$ & $0.0\%$ & $0.0\pm0.0$ & $0.0\%$\tabularnewline
\tableline
$300$ & $150$ & $1.9\pm1.4$ & $86.5\%$ & $6.5\pm2.5$ & $>99.9\%$ & $12.3\pm3.5$ & $>99.9\%$\tabularnewline
 & $200$ & $1.0\pm0.9$ & $63.8\%$ & $3.3\pm1.7$ & $97.5\%$ & $6.1\pm2.5$ & $99.8\%$\tabularnewline
 & $250$ & $0.5\pm0.7$ & $38.4\%$ & $1.5\pm1.2$ & $79.1\%$ & $2.8\pm1.7$ & $93.0\%$\tabularnewline
 & $300$ & $0.1\pm0.4$ & $12.7\%$ & $0.4\pm0.6$ & $33.7\%$ & $0.7\pm0.8$ & $51.5\%$\tabularnewline
 & $350$ & $0.1\pm0.3$ & $7.7\%$ & $0.2\pm0.5$ & $19.1\%$ & $0.4\pm0.6$ & $30.4\%$\tabularnewline
 & $400$ & $0.0\pm0.2$ & $3.7\%$ & $0.1\pm0.3$ & $10.1\%$ & $0.2\pm0.4$ & $15.4\%$\tabularnewline
 & $450$ & $0.0\pm0.1$ & $1.5\%$ & $0.0\pm0.2$ & $3.3\%$ & $0.1\pm0.2$ & $5.5\%$\tabularnewline
 & $500$ & $0.0\pm0.1$ & $0.4\%$ & $0.0\pm0.1$ & $0.7\%$ & $0.0\pm0.1$ & $0.8\%$\tabularnewline
 & $550$ & $0.0\pm0.0$ & $0.0\%$ & $0.0\pm0.0$ & $0.0\%$ & $0.0\pm0.0$ & $0.0\%$\tabularnewline
 & $600$ & $0.0\pm0.0$ & $0.0\%$ & $0.0\pm0.0$ & $0.0\%$ & $0.0\pm0.0$ & $0.0\%$\tabularnewline
\tableline
$250$ & $150$ & $1.6\pm1.3$ & $78.2\%$ & $5.5\pm2.5$ & $99.8\%$ & $10.6\pm3.3$ & $>99.9\%$\tabularnewline
 & $200$ & $0.7\pm0.8$ & $47.2\%$ & $2.4\pm1.6$ & $89.8\%$ & $4.4\pm2.1$ & $98.5\%$\tabularnewline
 & $250$ & $0.1\pm0.4$ & $12.7\%$ & $0.5\pm0.7$ & $39.6\%$ & $1.0\pm1.0$ & $61.6\%$\tabularnewline
 & $300$ & $0.1\pm0.2$ & $4.8\%$ & $0.2\pm0.5$ & $20.5\%$ & $0.4\pm0.7$ & $36.1\%$\tabularnewline
 & $350$ & $0.0\pm0.1$ & $1.6\%$ & $0.1\pm0.3$ & $7.9\%$ & $0.2\pm0.4$ & $15.4\%$\tabularnewline
 & $400$ & $0.0\pm0.1$ & $0.3\%$ & $0.0\pm0.1$ & $1.7\%$ & $0.0\pm0.2$ & $2.8\%$\tabularnewline
 & $450$ & $0.0\pm0.0$ & $0.0\%$ & $0.0\pm0.0$ & $0.0\%$ & $0.0\pm0.0$ & $0.0\%$\tabularnewline
 & $500$ & $0.0\pm0.0$ & $0.0\%$ & $0.0\pm0.0$ & $0.0\%$ & $0.0\pm0.0$ & $0.0\%$\tabularnewline
\tableline
$200$ & $150$ & $1.2\pm1.1$ & $69.2\%$ & $4.1\pm2.0$ & $98.4\%$ & $7.5\pm2.8$ & $>99.9\%$\tabularnewline
 & $200$ & $0.2\pm0.4$ & $20.1\%$ & $0.8\pm0.9$ & $52.9\%$ & $1.4\pm1.1$ & $76.3\%$\tabularnewline
 & $250$ & $0.1\pm0.3$ & $8.0\%$ & $0.3\pm0.5$ & $23.7\%$ & $0.5\pm0.7$ & $38.5\%$\tabularnewline
 & $300$ & $0.0\pm0.1$ & $1.8\%$ & $0.1\pm0.3$ & $6.7\%$ & $0.1\pm0.4$ & $10.6\%$\tabularnewline
 & $350$ & $0.0\pm0.0$ & $0.1\%$ & $0.0\pm0.0$ & $0.2\%$ & $0.0\pm0.0$ & $0.2\%$\tabularnewline
 & $400$ & $0.0\pm0.0$ & $0.0\%$ & $0.0\pm0.0$ & $0.0\%$ & $0.0\pm0.0$ & $0.0\%$\tabularnewline
\tableline
$150$ & $150$ & $0.3\pm0.5$ & $23.6\%$ & $1.1\pm1.0$ & $66.7\%$ & $2.0\pm1.4$ & $87.8\%$\tabularnewline
 & $200$ & $0.1\pm0.3$ & $7.2\%$ & $0.3\pm0.5$ & $24.9\%$ & $0.5\pm0.7$ & $39.6\%$\tabularnewline
 & $250$ & $0.0\pm0.1$ & $0.6\%$ & $0.0\pm0.1$ & $1.7\%$ & $0.0\pm0.2$ & $3.5\%$\tabularnewline
 & $300$ & $0.0\pm0.0$ & $0.0\%$ & $0.0\pm0.0$ & $0.0\%$ & $0.0\pm0.0$ & $0.0\%$\tabularnewline
\tableline
\end{tabular}
\tablecomments{As in Table~\ref{tab:upper-mass-limit-single-stars} but now it is assumed
that binary interactions took place in all massive binaries with initial orbital 
periods shorter than $5\,\mathrm{d}$ such that higher masses than the stellar
upper mass limit can be achieved because of binary mass transfer. Stars with masses
smaller than the upper mass limit are either (effectively) single stars or again
products of binary mass exchange in binaries with initial orbital periods shorter
than $5\,\mathrm{d}$.}
\end{table*}

\clearpage

\section{Star formation histories cont.}\label{sec:sfh-cont}
\begin{figure*}[H]
\center
\includegraphics[width=0.7\textwidth]{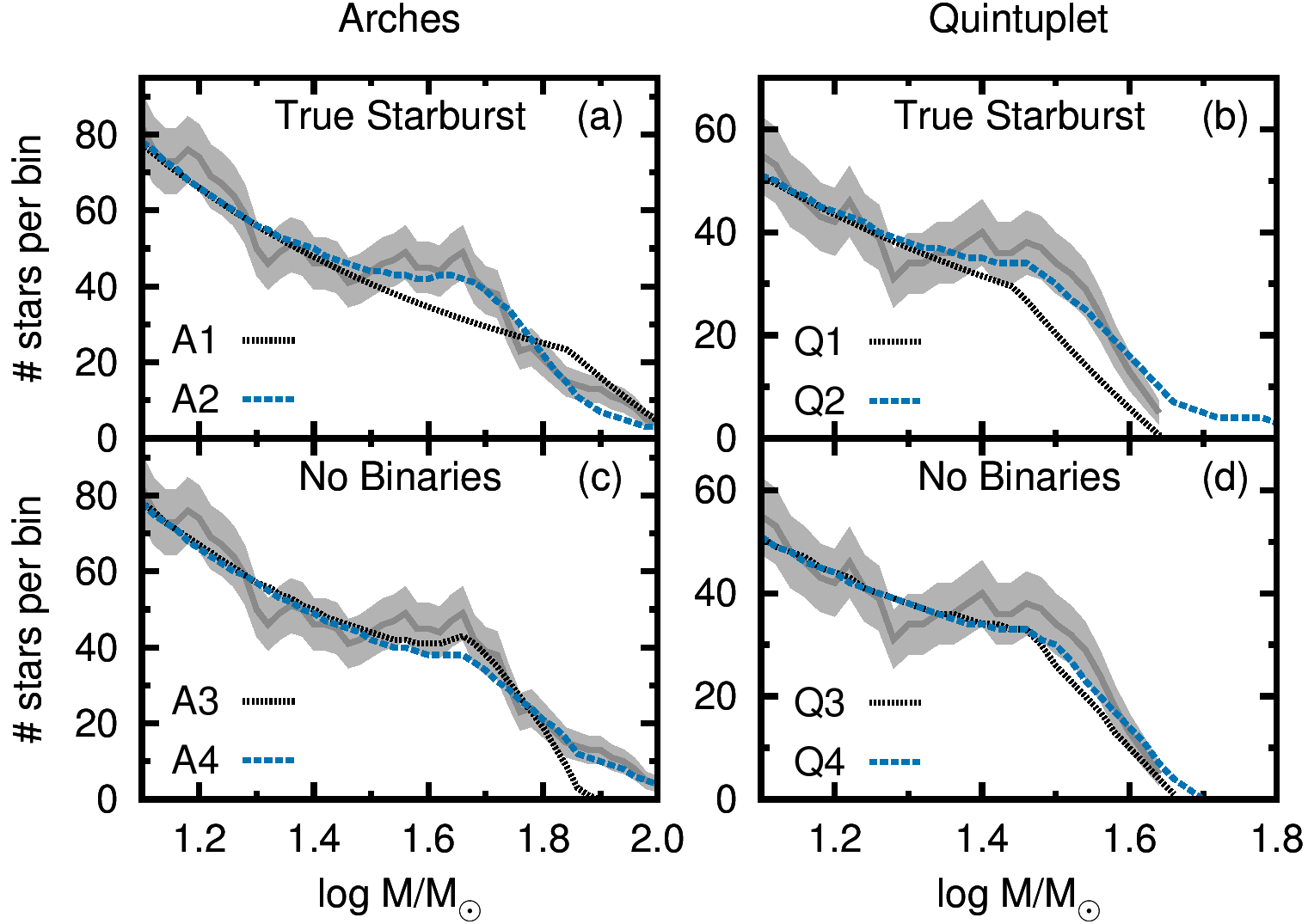}
\caption{Modelled mass functions using different star formation scenarios compared
to observations of the Arches and Quintuplet cluster. The left panels
(a) and (c) are for the Arches cluster whereas the right panels
(b) and (d) are for the Quintuplet cluster. In the top panels
(a) and (b), we compare the observations to our best fitting starburst models
including binary stars and to simple power-law mass functions truncated
at the observed maximum mass. In the bottom panels (c) and (d), we
show mass functions composed only of single stars with more complex
star formation scenarios. The individual star formation models A1-A4 and Q1-Q4 
are explained in Tab.~\ref{tab:sf-models} together with their least-square 
deviation $\chi^2$ (cf. Eq.~\ref{eq:least-squares}). All modelled mass functions 
are binned in the same way as the observations (see Sec.~\ref{sec:binning-procedure}).}
\label{fig:sfh}
\end{figure*}

Here, we investigate changes in the mass function due to
a star formation scenario which deviates from a true starburst in order to understand
whether the observed mass functions of Arches and Quintuplet can be reproduced without binaries.
We analyse two scenarios: (a) a period of prolonged but constant star formation rate
and (b) two instantaneous
starbursts separated in time. The latter scenario (b) not only represents a two 
stage starburst within one cluster but also two merged star clusters
where stars in each cluster are coeval. This situation most probably applies 
to the massive star cluster R136 in the Large Magellanic Cloud which is thought 
to be a double cluster in the process of merging \citep{2012ApJ...754L..37S}. 
We compute mass functions for the star
formation scenarios (a) and (b) (which include single, true starbursts) and
search for parameter values that minimise the least-square deviation,
of the modelled ($y_{{\rm model},i}$) from the observed ($y_{{\rm obs},i}$)
mass functions of the Arches and Quintuplet clusters
assuming Poisson uncertainties, i.e. $\sigma_{{\rm obs},i}^{2}=y_{{\rm obs},i}$,
\begin{equation}
\chi^{2}=\frac{1}{N}\sum_{i=1}^{N}\frac{\left(y_{{\rm model},i}-y_{{\rm obs},i}\right)^{2}}{\sigma_{{\rm obs},i}^{2}},\label{eq:least-squares}
\end{equation}
where $N$ is the number of mass bins. Exemplary star formation scenarios are
described in Table~\ref{tab:sf-models} and the resulting mass functions
are compared to observations in Fig.~\ref{fig:sfh}. Among these examples
are those star formation scenarios that lead to the best fits (models A2, A4, Q2 and Q4).

\begin{table*}[H]
\caption{Star formation (SF) scenarios used to compute the mass
functions which are compared to observations in Fig.~\ref{fig:sfh}.}
\label{tab:sf-models}
\centering
\begin{tabular}{ccccccp{9cm}}
\tableline 
SF model & $t_{1}/{\rm Myr}$ & $t_{2}/{\rm Myr}$ & $f_{{\rm B}}$ & $\chi_{{\rm tot}}^{2}$ & $\chi_{{\rm peak}}^{2}$ & Description\tabularnewline
\tableline
\tableline 
A1 & --- & --- & --- & $1.19$ & $1.59$ & power law mass function truncated at the most massive observed star;
power law index $\Gamma=-0.7$\tabularnewline
A2 & $3.5$ & --- & $100\%$ & $0.57$ & $0.36$ & single starburst at $t_{1}$\tabularnewline
A3 & $3.2$ & $3.3$ & $0\%$ & $1.85$ & $0.31$ & constant SF between $t_{1}$ and $t_{2}$\tabularnewline
A4 & $0.7$ & $3.3$ & $0\%$ & $0.49$ & $0.69$ & two starbursts at $t_{1}$ and $t_{2}$\tabularnewline
\tableline
Q1 & --- & --- & --- & $2.11$ & $4.23$ & power law mass function truncated at the most massive observed star;
power law index $\Gamma=-0.7$\tabularnewline
Q2 & $4.8$ & --- & $60\%$ & $0.56$ & $0.36$ & single starburst at $t_{1}$\tabularnewline
Q3 & $4.8$ & --- & $0\%$ & $0.70$ & $0.97$ & single starburst at $t_{1}$\tabularnewline
Q4 & $4.5$ & $4.7$ & $0\%$ & $0.43$ & $0.46$ & constant SF between $t_{1}$ and $t_{2}$\tabularnewline
\tableline
\end{tabular}
\tablecomments{Given are the primordial binary fraction $f_{\rm B}$ for each model 
as well as the least-square deviation $\chi^2_{\rm tot}$ in the total mass range ($1.1\leq \log M/\msun \leq 2.0$)
and $\chi^2_{\rm peak}$ in a mass region around the wind-mass-loss peak 
($1.4\leq \log M/\msun \leq 1.8$ for Arches and $1.3\leq \log M/\msun \leq 1.6$ for Quintuplet).
The best fit models are A2, A4, Q2 and Q4.}
\end{table*}

From Table~\ref{tab:sf-models} and the top panels of Fig.~\ref{fig:sfh} 
it is evident that simple power-law mass functions (models A1 and Q1) 
fit the observed mass functions of Arches and Quintuplet
much worse than the best single starburst models including binary
stars (A2 and Q2). Especially the mass region around the wind-mass-loss peak
is not fitted well by models A1 and Q1 (see $\chi^2_{\rm peak}$ in Table~\ref{tab:sf-models}). 
The mass functions of the Arches and Quintuplet clusters do not follow simple power laws.

In the bottom panels of Fig.~\ref{fig:sfh}, we also present
models of the observed mass functions of the Arches and Quintuplet clusters \emph{without}
binary stars. We do not find satisfactory models which fit peak and tail simultaneously 
with a single starburst without binaries.
Model A3 for example fits the peak due to mass-loss well ($\chi^2_{\rm peak}=0.31$) but
fails to explain the high mass end of the observed mass function ($\chi^2_{\rm tot}=1.85$).

We can improve this situation by adding a younger stellar population that fits the tail.
Such a scenario is given by model A4 which fits the peak and the total mass function. 
However, this two-component model has an age spread of $2.6\,{\rm Myr}$ which
is more than twice as large as the observed age discrepancy of about $1\,{\rm Myr}$
in the Arches cluster \citep{2008A&A...478..219M} and much larger
than the expected star formation duration.

A two component solution is not needed to model the observed mass function of the Quintuplet cluster
because the tail of the mass function is not very pronounced. Consequently
our models Q3 and Q4 predict no or a too small age spread --- contrary to the observations.

The age spread of $0.2\,\mathrm{Myr}$ of model Q4 might be compatible with the above estimated star 
formation duration given the quite uncertain core radius and velocity dispersion of Quintuplet.
The single starburst model Q3 is shown to illustrate the 
difference between the mass functions with (Q2) and without (Q3) binaries.
The tail of the mass function is however underestimated in
the observed mass function in Fig.~\ref{fig:sfh} because no self-consistent mass determination
for the three WNh stars in the core of Quintuplet is available \citep{2012A&A...540A..57H}.
If the tail were visible, we could of course model it by an additional
younger population as is in Arches. 

In summary, we conclude that we can reproduce the mass functions
of Arches and Quintuplet without binaries but with freedom in the
star formation history. However, the best fit star formation parameters (e.g. 
the age spread of $2.6\,{\rm Myr}$) are inconsistent with other observables.
Our single starburst models which include binaries are thus the only models
which fulfil all observational constraints. These models are also consistent
with a star formation duration of the order of the crossing time of
the cluster. 

\clearpage

\bibliographystyle{apj}

\end{document}